\documentclass[acmsmall]{acmart}
\AtBeginDocument{%
  }

\setcopyright{acmlicensed}
\acmJournal{PACMHCI}
\acmYear{2025} \acmVolume{9} \acmNumber{2} \acmArticle{CSCW168} \acmMonth{4}\acmDOI{10.1145/3711066}

\graphicspath{{figs/}{figures/}{pictures/}{images/}{./}} 
\usepackage{cleveref}
\usepackage{tabu}                      
\usepackage{booktabs}                  
\usepackage{lipsum}                    
\usepackage{mwe}                       
\usepackage{mathptmx}
\usepackage{color}
\usepackage{fancyvrb}
\usepackage{xcolor}
\usepackage{array}
\usepackage{enumitem}

\newcommand{\tocheck}[1]{\textcolor{black}{#1}}
\usepackage{xspace,xpunctuate}

\newcommand{\ie}{\textit{i.e.},\xspace}
\newcommand{\etal}{\xspace\textit{et al.}\xspace}
\newcommand{\eg}{\textit{e.g.},\xspace}
\newcommand{\re}[1]{\textcolor{black}{#1}}
\newcommand{\minor}[1]{\textcolor{black}{#1}}

\begin{document}

\title{SimSpark: Interactive Simulation of Social Media Behaviors}

\author{Ziyue Lin}
\email{ziyuelin917@gmail.com}
\orcid{0009-0002-5485-7379}
\affiliation{%
  \institution{School of Data Science, Fudan University}
  \city{Shanghai}
  \country{China}
}

\author{Yi Shan}
\email{20302010026@fudan.edu.cn}
\orcid{0009-0006-8026-0735}
\affiliation{%
  \institution{School of Data Science, Fudan University}
  \city{Shanghai}
  \country{China}
}
\author{Lin Gao}
\email{lgao.lynne@gmail.com}
\orcid{0009-0004-1613-1774}
\affiliation{%
  \institution{School of Data Science, Fudan University}
  \city{Shanghai}
  \country{China}
}
\author{Xinghua Jia}
\email{18300290007@fudan.edu.cn}
\orcid{0009-0001-6028-387X}
\affiliation{%
  \institution{School of Data Science, Fudan University}
  \city{Shanghai}
  \country{China}
}
\author{Siming Chen}
\authornote{Siming Chen is the corresponding author.}
\email{simingchen@fudan.edu.cn}
\orcid{0000-0002-2690-3588}
\affiliation{%
  \institution{School of Data Science, Fudan University}
  \city{Shanghai}
  \country{China}
}
\affiliation{%
  \institution{Shanghai Key Laboratory of Data Science}
   \streetaddress{1 Th{\o}rv{\"a}ld Circle}
  \city{Shanghai}
  \country{China}
}

\renewcommand{\shortauthors}{Ziyue Lin, Yi Shan, Lin Gao, Xinghua Jia, and Siming Chen}
\begin{abstract}
  Understanding user behaviors on social media has garnered significant scholarly attention, enhancing our comprehension of how virtual platforms impact society and empowering decision-makers.
  Simulating social media behaviors provides a robust tool for capturing the patterns of social media behaviors, testing hypotheses, and predicting the effects of various interventions, ultimately contributing to a deeper understanding of social media environments.
  Moreover, it can overcome difficulties associated with utilizing real data for analysis, such as data accessibility issues, ethical concerns, and the complexity of processing large and heterogeneous datasets.
  However, researchers and stakeholders need more flexible platforms to investigate different user behaviors by simulating different scenarios and characters, which is not possible yet.  
  Therefore, this paper introduces SimSpark, an interactive system including simulation algorithms and interactive visual interfaces which is capable of creating small simulated social media platforms with customizable characters and social environments. 
  We address three key challenges: generating believable behaviors, validating simulation results, and supporting interactive control for generation and results analysis. 
  A simulation workflow is introduced to generate believable behaviors of agents by utilizing large language models. 
  A visual interface enables real-time parameter adjustment and process monitoring for customizing generation settings. 
  A set of visualizations and interactions are also designed to display the models' outputs for further analysis. 
  Effectiveness is evaluated through case studies, quantitative simulation model assessments, and expert interviews.
\end{abstract}



\begin{teaserfigure}
  \includegraphics[width=\textwidth]{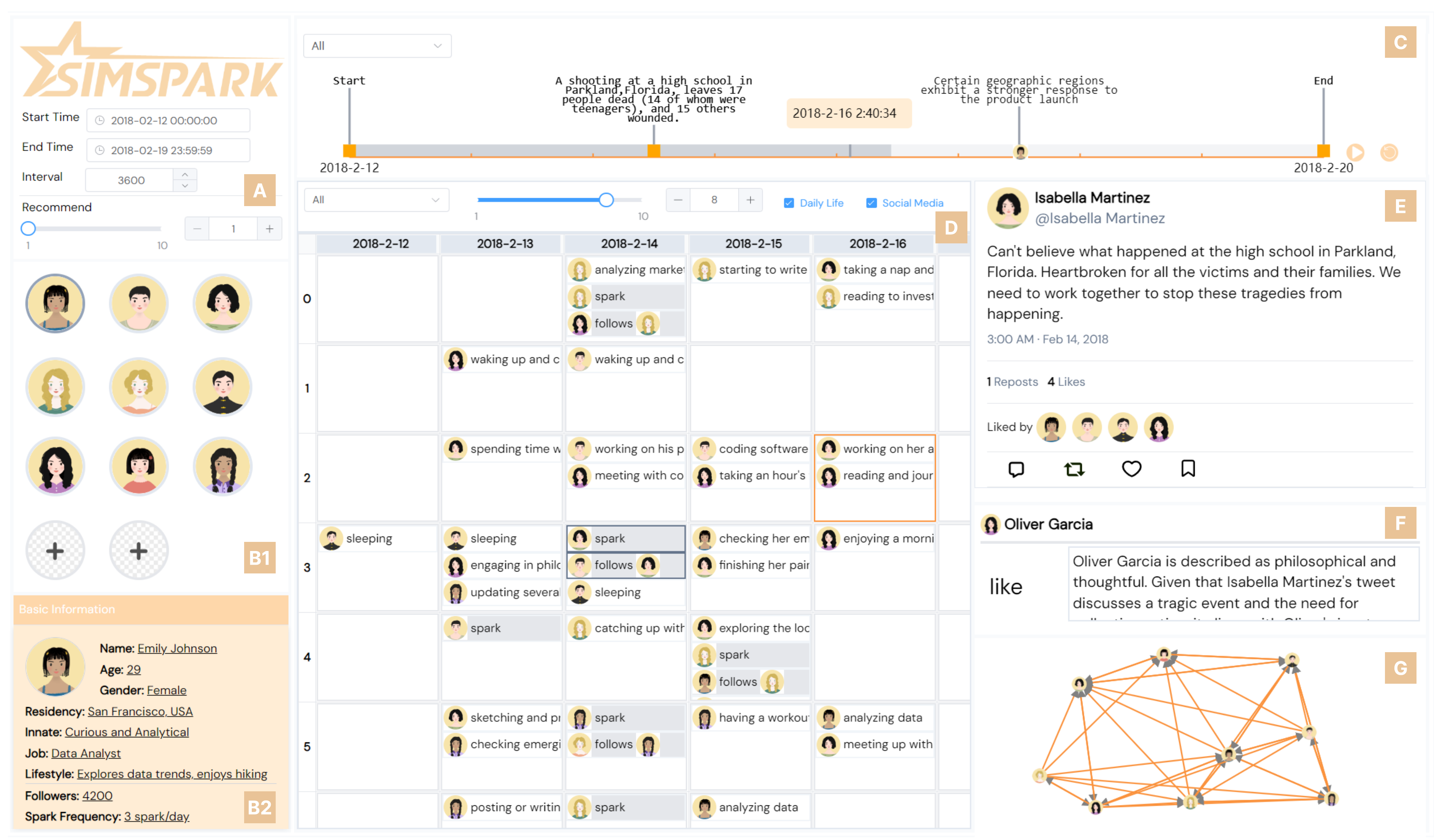}
  \caption{The interface of SimSpark.~(A) Setting Panel allows users to configure environmental parameters.~(B1) Avatar Panel displays agents' avatars.~(B2) Basic Information Panel allows users to configure demographic information and social habits of selected agents.~(C) Timeline shows the public events and simulation process.~(D) Calender View displays agents' behaviors.~(E) Sparkle View shows simulated social media.~(F) Reasoning View displays the reason behind simulated social media behaviors of the agent.~(G) Follow Network and Hidden Reasoning can be switched in the view. Follow Network illustrates the following relationships of agents. Hidden Reasoning displays the reason why agents don't act, which is shown in~\Cref{fig:casexxx}B.}
  \label{fig:teaser}
\end{teaserfigure}



\received{January 2024}
\received[revised]{July 2024}
\received[accepted]{October 2024}

\maketitle
\section{Introduction}
Social media technology has reshaped how we connect with others, share information, and engage with the world in today's digital age~\cite{KAPLAN201059,doi:10.1080/15456870.2015.972282}. 
Individuals can nearly access social platforms anytime due to the advent of smartphones and mobile apps~\cite{kapoor_advances_2018}.
Social media has become an integral part of our daily lives.
Users are allowed to create and share their own experiences, perspectives, and reflections on social media.
In addition, connecting with friends and forging new relationships through social media transcend geographical boundaries.
The democratization of content creation has fostered a sense of individuality and diversity online, which has received significant attention from professionals in a wide range of fields~\cite{kapoor_advances_2018}.
Social media user behaviors research aims to obtain a deeper understanding of how virtual platforms influence our lives and society, which further empowers marketers, policymakers, and researchers to make informed decisions.
For example, Lundmark\etal~\cite{lundmark_little_2017} highlighted the emerging impact of social media in the legalization process by indicating that the degree to which a company employs Twitter is linked to consistently elevated levels of Initial Public Offering~(IPO) underpricing. 
These studies, encompassing various interdisciplinary aspects, hold significant importance in our interconnected world.

\minor{Obtaining data from real social media platforms is challenging.
Social media APIs often impose rate limits to control the volume of data accessed within a given time frame and charge significant fees for advanced features, broader data scopes, or higher rate limits.
For example, \textbf{X} provides access to essential API endpoints, allowing users to collect up to 10,000 tweets per month at a cost of approximately \$100 a month~\cite{twitter}, increasing in the past few years.}
\minor{Second,} utilizing real social media data for research presents several significant difficulties, which stem from the inherently noisy nature of the data itself, the ethical and legal considerations, and the technical complexities of processing large and heterogeneous datasets.
Social media data is often noisy and contains irrelevant information, spam, or bots, which can distort analysis results.
It is challenging to filter and clean the data\minor{~\cite{GHANI2019417}}.
\minor{Thrid,} the ethical quandaries surrounding the utilization of genuine data in research contexts cannot be disregarded\minor{~\cite{doi:10.1089/cyber.2012.0334,10.1007/s10676-010-9227-5}}.  
One of the most significant concerns pertains to the debate over whether such social media data should be categorized as public or private information. 
Although most social platforms have terms and conditions~(T\&Cs) which frequently contain clauses regarding third-party access to user data, Boyd\etal~\cite{doi:10.1080/1369118X.2012.678878} stated that the process of assessing research ethics could not be ignored simply because the data appears to be public. 
Researchers also cannot ensure that participants are adequately informed and have provided their consent when conducting experiments on real social platforms to discover user behavior patterns.
\re{Additionally, extracting the data needed to address specific research questions from raw datasets requires significant computational resources and a multidisciplinary approach to ensure the reliability and validity of the extracted data\minor{~\cite{SOBKOWICZ2012470,10.3389/fpsyg.2014.00668}}.
On the other hand, }
social media researchers and stakeholders are somewhat constrained in their capacity to explore diverse user behaviors by analyzing real social media data, particularly when attempting to \re{test hypotheses about social media behavior in a controlled, replicable environment and predict the outcomes of various interventions}. 
This limitation presents a significant research gap, which remains elusive. 
For instance, imagine a situation where researchers aim to investigate how various consumer segments react to a new product launch on social media. 
\re{Moreover, what type of promotional content will attract more attention from consumers is also a question worth exploring.}
In the absence of flexible \re{research tool}, accurately replicating and examining these distinct consumer responses becomes an intricate and unmanageable task.

Motivated by the limitations mentioned above, we propose an interactive method \re{based on agent-based modeling~(ABM)} for supporting simulating social media behavior, which allows \re{social media content creators and user behavior researchers} to conduct controlled simulation where various parameters, scenarios, and user behaviors can be manipulated and studied systematically. 
There are three major challenges to achieving the objectives.
First, generating believable social media user behavior is of great importance.
Nevertheless, generating behavior that convincingly mimics genuine human actions is complex. 
It requires a deep understanding of human behavior on social media, including language use, posting habits, and interaction dynamics.
Existing social bot techniques have primarily focused on content generation, often producing low-quality and repetitive content that is highly susceptible to detection as non-authentic users~\cite{9014365,10.1145/3459637.3481949}.
Moreover, those bots typically exhibit unusual behaviors, such as excessive posting or commenting, and lack the ability to bond with other users to form natural social network patterns.
With the rise of \re{large language models~(LLMs)}, many researchers utilize these models to simulate human behavior~\cite{10.1145/3526113.3545616,park2023generative}. 
However, there is a lack of tailored methodologies for simulating diverse social media behaviors.
Therefore, designing an adequate simulation model to generate believable social media behavior is still challenging.
The second challenge is validating the output of the simulation model. 
On the one hand, many social media behavior models, particularly those based on deep learning and neural networks, are difficult to interpret, making it unclear why certain decisions or behaviors are generated.
It poses a significant challenge for researchers in determining whether unusual generated behaviors are caused by inherent complexities of human behaviors or unsatisfactory model performance.
On the other hand, assessing simulated behaviors requires considering various dimensions and ensuring a more nuanced and robust evaluation of the subject under scrutiny, making it difficult to interpret and verify the model outputs. 
The third challenge is further supporting interactive control for data generation and result analysis.
Many generation models require complicated parameterization involving character profiles and environment settings.
What's more, selecting appropriate parameter values can be subjective, requiring continuous parameter adjustment to achieve a satisfactory outcome eventually. 
Most social media behavior researchers come from humanities backgrounds and lack proficient programming skills, which can pose significant challenges for debugging models.
Hence, facilitating parameter configuration for researchers proves to be an imperative task.
Furthermore, the diversity of behavioral data leads to high heterogeneity, making it challenging to extract meaningful insights from generated outputs.

\minor{To answer the research problem of how human-computer interaction can enable more flexible and customizable simulations of social media behaviors to better align with the research needs of target users,} we develop ${SimSpark}$, a novel interactive system that supports creating a simulated \re{text-only} social platform containing multiple generative characters, which enables users to customize characters and control social environments. 
To address the first challenge, we adopt intelligent agent techniques powered by LLMs to simulate social media behaviors.
\re{An agent can perceive its environment through sensors and act upon that environment using effectors~\cite{russell2016artificial}.}
Conducting a literature review about social bot detection algorithms~\cite{feng_twibot-22_2022,10.1145/3409116}, we design a simulation workflow including system configuration, social media engine, and agent cognitive architecture. 
Existing detection algorithms are designed to recognize patterns of behavior that resemble automated or scripted actions. 
These patterns can be analyzed to understand what constitutes inauthentic behavior. 
We utilize this knowledge to create social media agents that exhibit genuine, human-like behavior.
What's more, we leverage LLMs to power the simulation process for more believable outputs, especially human-like post content.
For the second challenge, LLMs' ability to generate coherent and contextually relevant text can provide agents' reasoning for their actions while generating behaviors to enhance the interpretability of the model.
Furthermore, we design a visual interface that includes a set of visualizations and interactions to reveal behavior patterns of generative agents and facilitate more intuitive comprehension of patterns.
For the third challenge, the visual interface also enables real-time parameter adjustment and process monitoring for customizing generation settings, which does not necessitate users to possess programming skills.
In particular, not only do we visualize the simulated behaviors and ``following'' network of agents, but we also display the reasoning of deciding to act or abstain for further analysis.
Finally, we evaluate the effectiveness of our system by conducting case studies in different scenarios, a quantitative experiment of the model performance, and interviewing domain experts for user feedback.

In summary, our main contributions are as follows:
\begin{itemize}
\item We design a simulation workflow to generate believable social media behaviors of agents by extending a large language model, which contains three components: system configuration, social media engine, and cognitive architecture. 
\re{It provides an innovative social computing method for studying social media dynamics and supports diverse research and practical applications.}
\item We propose a novel interactive system called ${SimSpark}$ for users to create small simulated social platforms with customizable agents and social environments.
\re{The flexibility of the system enables users to simulate a wide range of social media scenarios, tailoring agents and environments to fit specific research questions or application needs.
Enhanced interactivity improves the usability of the system, making it accessible to a broader range of users, including those without advanced technical skills.}
\item We evaluate the effectiveness of our system through case studies, a quantitative experiment, and expert interviews. 
\re{The evaluations provide positive feedback, serving as valuable references for future research and development.}
\end{itemize}
\section{Related Work}
This section discusses relevant studies about social media behavior simulation, including intelligent agents, agents powered by large language models, and agent-based modeling\re{~(ABM)}. 
Intelligent agents exhibit certain levels of intelligence, capable of perceiving their environment, making decisions, and executing actions to achieve predefined goals or objectives. 
Recently, researchers have been using large language models to create agents, capitalizing on their remarkable emergent capabilities and widespread popularity. 
Agent-based modeling employs multiple and heterogeneous agents to model and simulate complex systems or phenomena in a social environment. 
When applied to model social media, ABM allows the simulation of various aspects of social media platforms, user behaviors, and interactions.

\subsection{Intelligent Agents Development}
The study of intelligent agents has a long history in artificial intelligence research. 
Intelligent agents exhibit certain levels of intelligence, capable of perceiving their environment, making decisions, and executing actions to achieve predefined goals or objectives~\cite{10.1007/BFb0013570,wooldridge_jennings_1995}.
Like vivid characters from movies, books, animation, or theater, believable agents are characterized by their capacity to simulate human-like behaviors, emotions, and thinking processes. 
Benefiting from autonomous algorithms, users can interact with such powerful, personality-rich agents~\cite{bates1994role,laird_human-level_2001}.
These intelligent agents are well known for appearing in video games as non-player characters~(NPCs), contributing to more authentic and captivating user experiences~\cite{riedl_interactive_2021}. 
They can create a sense of a living, breathing world within the game, where characters exhibit realistic behaviors, react to changing circumstances, and even learn from their interactions with players~\cite{brenner_creating_2010,isbister_consistency_2000}. 
Due to the excessively vast state space, the developments of the most believable agents tend to simplify either the environmental rules or behavioral guidelines to improve the feasibility of algorithms~\cite{10.1007/3-540-48834-0_5}.
When the environmental conditions are complex and agents have common behavioral traits, behavior trees, models of plan execution, are appropriate because of their maintainability, scalability, and extensibility~\cite{marcotte_behavior_2017}.
However, they cannot adapt to new or unexpected inputs or questions that fall outside their rule set since they rely on predefined behavior patterns~\cite{siu_evaluation_2021}. 
To seek more versatile and intelligent agents, developers have turned to machine-learning approaches, such as reinforcement learning, to overcome these inherent limitations~\cite{vinyals_grandmaster_2019}.
It requires a carefully crafted reward structure that reflects the quality of interactions, which can be subjective and context-dependent~\cite{8022767}.
By incorporating principles from cognitive psychology and neuroscience, developers can create agents that emulate human-like thought processes and decision-making, resulting in more realistic and believable behavior~\cite{10.1145/375735.376343}.
Cognitive models provide insights into how humans think, reason, and make decisions~\cite{holyoak2005cambridge}.
However, developing cognitive models often involves intricate algorithms and simulations of various cognitive processes, such as perception, memory, reasoning, and decision-making. 
Implementing and fine-tuning these models can be technically demanding.
Powered by large language models, Park\etal~\cite{park2023generative} proposed an effective cognitive architecture that supports forming long-term plans, and creating high-level reflections to simulate the two-day lives of 25 agents in ``Smallville''.

To simulate believable social media behaviors, we adapted a hierarchy of thinking kinds~\cite{rips_folk_1989} and the cognitive architecture proposed by Park\etal~\cite{park2023generative} with some specific modifications. 
In addition, we leveraged the excellent performance of large language models to improve the effectiveness of cognitive models.

\subsection{Agents Empowered by Large Language Models}
Large language models~(LLMs) offer the ability to understand and generate human-like language sophisticatedly at an unprecedented scale~\cite{bommasani2022opportunities,brown_language_2020}.
What's more, these models, such as ChatGPT~\cite{brown_language_2020}, are designed to engage in multi-turn dialogues, sustaining conversations over multiple exchanges and dynamically adapting their responses as the discussion evolves.
Thus, the interactivity of LLMs is a pivotal aspect of their utility and appeal~\cite{10.1145/3491101.3519729}.
\re{The potential of LLMs extends beyond mere text generation, as they can be harnessed to create agents that exhibit higher levels of intelligence and complexity in their behavior.
LLMs possess advanced language comprehension skills, enabling them to understand context, nuances, and subtle cues in text.
This capability allows agents to interpret and respond to complex environments accurately by detailed text descriptions.
By leveraging the extensive data on which LLMs are trained, agents can exhibit various behaviors, reflecting different characteristics, making the simulated interactions more realistic and varied.
Moreover, LLMs can be integrated with cognitive architectures that model human thought processes. This integration allows agents to exhibit higher-order cognitive functions, such as reasoning, reflecting, and planning~\cite{park2023generative}.
Therefore, with meticulously crafted prompt and simulation workflow, agents empowered by LLMs can be more adaptive, realistic, and intelligent.}
LLM-empowered agents leverage sophisticated language understanding and generation capabilities and can be generally classified into goal-oriented and simulation-oriented agents~\cite{xi2023rise}.
The goal-oriented agents are characterized by their emphasis on achieving predefined objectives or targets. 
LLMs agents are proven to be capable of playing open-ended creative games like Minecraft,  Dungeons and Dragons~\cite{callisonburch2022dungeons,wang2023voyager}. 
They are also used for the automatic creation of interactive stories for games~\cite{10.1145/3402942.3409599} based on their powerful ability of language generation.
On the other hand, simulation-oriented agents are designed to replicate and mimic real-world phenomena within a simulated environment.
Horton\etal~\cite{NBERw31122} replicated four classic experiments in economics by using LLMs to endow agents with various social preferences.  
GPTeach~\cite{markel2023gpteach}, an interactive chat-driven teacher training tool, allows novice educators to engage in practice sessions with simulated students.
Generating open-ended questionnaire responses for Human-Computer Interaction~(HCI) research by LLMs is a direction worth exploring~\cite{10.1145/3544548.3580688}.
According to users' community design, Social Simulacra~\cite{10.1145/3526113.3545616} used LLMs to generate social interactions interactively, encompassing posts, replies, and anti-social behaviors. 
$S^3$~\cite{gao2023s3} modeled the dynamic progression of public sentiment on social platforms in reaction to trending societal events.

Our framework falls into the classification of simulation-oriented agents.
Although Social Simulacra~\cite{10.1145/3526113.3545616} has enabled users to explore simulated group chatting in different scenarios by changing the community design, it implemented limited simulation configuration and social behaviors and lacked in-depth exploration of the underlying logic behind the behavior. 
We design a novel simulation workflow, including system configuration, social media engine, and agent cognitive architecture, especially to mimic human-like social media behaviors, such as posting, following, and replying.
In addition, a series of chain-of-thought prompts are designed to perform the reasoning process of choosing to undertake or abstain from an action, which improves the interpretability of model results and helps users validate simulation results~\cite{NEURIPS2022_9d560961}.

\subsection{Agent-based Modeling and Social Media}
Agent-based Modeling~(ABM) is a powerful technique simulating complex social systems, which involves creating a computational environment where autonomous agents interact with each other and their environment, following predefined rules or behaviors~\cite{1574234,5429318,helbing_agent-based_2012,doi:10.1073/pnas.072081299}. 
ABM enables researchers to explore emergent phenomena resulting from interactions among individual agents, providing insights into the collective behavior of the modeled system. These emergent phenomena can be observed at a macro level, revealing patterns, dynamics, or characteristics that may not be apparent when solely focusing on individual agents~\cite{1597399,savaglio_agent-based_2020,10.1007/978-3-642-54783-6_2}.
The approach is highly versatile and has been applied across diverse fields. 
For example, ABM stands as a robust simulation paradigm to effectively address challenges and provide comprehensive support for developing the Internet of Things ecosystem~\cite{savaglio_agent-based_2020}.
Covasim~\cite{kerr_covasim_2021}, an agent-based method, aims to help project epidemic trends, explore intervention scenarios, and estimate resource needs.
Axtell\etal~\cite{axtell2022agent} demonstrated that ABM had expanded our comprehension of various domains, including markets, industrial organization, labor economics, macroeconomics, and policy analysis.

ABM in social media modeling enables the examination of various phenomena, such as the spread of information or trends, the formation of social networks, and the impact of user behaviors on content dissemination.
Employing the ABS framework, the propagation of rumors can be examined by engaging in an internalized and resource-intensive process of networked interactions to measure its impact~\cite{7225278,7408553}. 
\re{Butler\etal~\cite{butler_misinformation_2024} introduced The Misinformation Game, an open-source, customizable online platform that simulates social media to study misinformation processing.}
Onuchowska\etal~\cite{onuchowska2019using} assessed the efficacy of various social media policy regulations in curbing malicious behavior through the ABM platform. 
ABM has also been applied to predict human behavior in social media concerning different topics and sentiments by simulating interactive behavior~\cite{doi:10.1177/0037549713477682,10.1007/978-3-642-54783-6_2}.
\re{Lyfe Agents~\cite{kaiya2023lyfeagentsgenerativeagents}, combining low-cost real-time responsiveness, leverage an option-action framework, asynchronous self-monitoring, and a Summarize-and-Forget memory mechanism to simulate intricate social behaviors in virtual societies.
Li\etal~\cite{li2023masqueradeexploringbehaviorimpact} studied on a Twitter-like network of LLM-driven bots, revealing enhanced individual camouflage, collective behavior patterns, and influential toxic behaviors.}
Törnberg\etal~\cite{törnberg2023simulating} leveraged LLMs and ABM to simulate users' social media behaviors and study how different news feed algorithms shape the quality of online conversations. 
\re{Although existing works support simple interactions, such as initializing agents~\cite{lin2023agentsimsopensourcesandboxlarge} or giving action commands to agents~\cite{nakano2022webgptbrowserassistedquestionansweringhuman,hong2023metagptmetaprogrammingmultiagent}, user involvement with the model is not sufficiently in-depth.}
In our paper, the design of interactive interfaces facilitates users customizing the characteristics of intelligent agents and social environments, \re{controlling the process of the simulation, real-time monitoring simulation,} and allowing for an intuitive review of model results for further analysis.

\re{\subsection{Research Gap and Objectives}
Although existing work has made significant progress in simulating social media behaviors by leveraging LLMs, \minor{such as Social Simulacra focusing on posts}~\cite{10.1145/3526113.3545616}, \minor{few works have} integrated the study of diverse social media behaviors, such as posting, liking, following, and replying. 
We tailored the simulation workflow for multiple social media behaviors based on human cognitive models and social media habits.
Second, the majority of current research~\minor{(\eg AgentSims~\cite{lin2023agentsimsopensourcesandboxlarge}, $S^3$~\cite{gao2023s3} )} has not thoroughly investigated human-computer interactivity, particularly how users interact dynamically with automated agent-based modeling techniques. 
This gap includes how users define agents and environments of simulation, how users understand the model outputs, and how users respond to simulated behaviors in real-time.
Thus, our research question focuses on exploring how human-computer interaction can support more flexible and customizable social media behavior simulations to better meet the research needs of target users.
In this paper, we develop a user-friendly interface that supports setting agents and environment by natural language command.
Additionally, users can monitor, pause, or reset the simulation process at any time to control the model precisely.
Furthermore, a set of visualizations is designed to uncover behavior patterns of generative agents and enable a more intuitive understanding of these patterns.}
\re{\section{Formative Study}
We conducted a formative study to derive the design requirements of our interactive system. 
As mentioned above, our target users are social media content creators and user behavior researchers.
We interviewed professionals from different backgrounds who were involved in studying social media data and user behavior.
\subsection{Interviewees}
We recruited \minor{five} stakeholders who are our target users to conduct an in-person semi-structured interview, respectively. 
Table \ref{tab:demogra} shows their detailed demographic information.
The sample consists of five individuals, including four females and one male, with ages ranging from 24 to 30 years old~($Age_{mean} = 26$, $Age_{std} = 2.34$).
The interviewees from academia and industry backgrounds hold various occupations, such as advertising, individual media, marketing, and postgraduate research. 
Their years of employment in their respective fields are all more than 2 years~($Employment_{mean} = 3.8$, $Employment_{std} = 2.49$).
The majority of them engage in studying social media data every day. 
The diversity provides a comprehensive understanding of different perspectives related to social media data and user behavior analysis.
Besides, each interview lasted more than one hour, and the interviewees were paid \textdollar50 an hour.
\subsection{Procedure}
Before the interview, we obtained the consent of the interviewees to record the meeting and use their demographic information for further analysis.
The semi-structured interview study was divided into three parts.
In the first part of the interview, we asked the interviewees about the nature of their work and the role that social media data analysis plays in their professional activities.
For example, we asked them \textit{``Can you briefly describe your background and current role?''} and \textit{``Why is it necessary to analyze social media data in your work?''}
In the second part, we focus on the difficulties encountered by the interviewees when analyzing data.
We asked them the question \textit{ ``What are some of the main challenges you face when analyzing social media data?''}
In the third part, we introduced our idea of developing an interactive system to simulate virtual social media data.
We inquired whether such a system would be beneficial for their work and what specific issues it could help them address.
In addition, the interviewees also provided suggestions for the system.
After completing all the interviews, we organized the feedback and iteratively summarized it into findings based on the recordings.
\subsection{Findings}
Based on our recordings, we summarized three findings, including the data dilemma, the characteristics of social media, and the design of the simulation system.
We aimed to derive system requirements and guide system design through these findings.
\subsubsection{Data Dilemma.}
All interviewees agreed that utilizing real social media data for research involves numerous difficulties.
First, social media platforms often restrict access to their data, providing limited APIs or imposing strict usage limits. 
P1 said \textit{``It's really hard to get data from social media platforms because they have a lot of restrictions.''}
Second, cleaning and processing data is very cumbersome, especially for researchers without a background in data analysis and programming.
Ensuring data is anonymized to protect user privacy while retaining its utility for analysis is a complex and often imperfect process.
Processing raw datasets into structured data based on research requirements for further analysis, such as handling noise, cleaning irrelevant data, and standardizing formats, is also very challenging. 
P4 and P5 mentioned that many research projects stagnate due to the lack of high-quality datasets, and researchers often lack the skills for web scraping or data processing.}

\re{After our idea informed them, they all showed great interest in our research.
By defining agents and environments based on research needs, they can quickly obtain required and relevant data without a cleaning process.
Furthermore, P1, P2, and P4 mentioned that simulation could enhance hypothesis testing by providing a controlled, flexible, and replicable environment, especially where testing in real-world settings would be difficult or impossible.
P2 said \textit{``I can use the simulation tools to test the feedback my created content would receive before its official release.''}
P4 also said \textit{``Simulating the spread of misinformation on social media would allow us to observe how false information propagates without the ethical and practical issues of conducting such an experiment in real life.''}
However, they also expressed some concerns about our approach.
The believability of simulated behaviors is crucial for the validity of research findings and the credibility of the study.
Believable results are more likely to be applicable to real-world scenarios, enhancing the generalizability of the research.
P3 said \textit{``If the simulation accurately reflects how users interact with content on social media, we can better predict engagement and optimize strategies for future campaigns.
Otherwise, the insights we gain might be misleading and fail to inform our strategies and decisions effectively.''} 
\subsubsection{Characteristics of Social Media.}
We concluded some key characteristics of social media that greatly influence the research of social media from interviewees. 
These were very insightful for our method design.
First, recommendation systems play a significant role in shaping the landscape of social media research, which personalize content for individual users based on their preferences and past interactions. 
All interviewees agreed that the application of recommendation algorithms is becoming increasingly widespread. 
p4 said \textit{``This can create echo chambers or filter bubbles, where users only see content that reinforces their existing beliefs and interests.''}
Second, key opinion leaders~(KOLs)~\cite{scher2021key} have a significant impact on users through their perceived authority, trustworthiness, and ability to engage and connect with their audience.
These KOLs are distinguished by their expertise in specific fields, substantial follower base, and active engagement on social media platforms.
They play a pivotal role in shaping public opinion, conveying specific information, and sometimes are invited by brands to promote products.
KOLs create engaging content that resonates with their audience.
High engagement rates can amplify the reach of the content and further influence user behavior.
P3 said \textit{``For example, fashion enthusiasts follow popular fashion accounts on Instagram to stay updated with the latest trends, news, and releases from designers.''}
Third, social media networks have become an integral part of modern society, transforming the way people interact with each other.
Users tend to connect with others who have similar interests and opinions, which reinforces existing beliefs and limits exposure to diverse perspectives, further entrenching users in their viewpoints.
P5 said \textit{``Users are more easily influenced by the behaviors and opinions of their peers on social networks.''}
Social networks may cause users anxiety by showcasing others' experiences and achievements, and users may feel compelled to stay constantly updated.
\subsubsection{Design of Simulation System.}
All interviewees responded positively to our research.
They provided valuable suggestions and expressed their expectations about the interactive simulation tools.
First, convenient customization of the simulation model ensures that the system can be effectively used by researchers without a deep understanding of the underlying simulation algorithms.
Users should be provided with intuitive and flexible tools to easily define and modify the simulation configurations involving the characteristics of agents and environments.
A user-friendly interface can reduce the system's learning curve. 
By leveraging LLMs, users can quickly define and modify simulation configurations using natural language as model input.
P1, P3, and P5 pointed out that natural language descriptions can ensure that agent characteristics align closely with user expectations.
Second, the necessity of interpretability of simulation results has been emphasized by all interviewees.
The black-box nature of LLMs makes it difficult to understand the decision-making process within the algorithm.
P2 said \textit{``I can't see how an algorithm makes its decisions, which can lead to mistrust and hesitation to use the technology.''}
Thus, simulation behaviors should be presented clearly and comprehensibly to help users quickly grasp the outcomes and underlying dynamics of the simulation. 
P1 and P2 mentioned that appropriate interactive visualization design, such as charts and graphs, would be very helpful.
P5 also said \textit{``I expect this system to showcase the logic and reasoning behind the simulation behaviors.''}
Third, the system should support real-time process monitoring of simulation.
To reduce the time and computational cost, real-time monitoring keeps users informed and allows for quick action to mitigate potential problems.
Users will have options to pause and resume the simulation or adjust the parameters to refine the simulation dynamically every time an anomaly is detected.
Furthermore, P4 said \textit{``Real-time adjustments and control will let us try out different scenarios and settings conveniently.''}
\subsection{Design Requirements}
}
After interviewing \re{our target users mentioned above} who conduct social media behaviors analysis in their research and conducting literature review~\cite{10.1145/3526113.3545616,gao2023s3,1574234,5429318,helbing_agent-based_2012,doi:10.1073/pnas.072081299, leong_ict-enabled_2015,thoma_establishing_2018}, we eventually specified four system design requirements:
\begin{enumerate}[label=\textbf{R{\arabic*}}, nolistsep]
\item 
\textbf{Simulate believable social media behaviors.} Believable is of paramount importance, serving as the foundation for subsequent research endeavors. Therefore, in addition to employing LLMs, it is imperative to propose a tailored simulation workflow specifically designed for social media behaviors.
\item 
\textbf{Facilitate simulation configuration.} The system should be designed to facilitate simulation configuration for researchers without a programming background. Additionally, it should also support real-time parameter adjustment and process monitoring.

\item 
\textbf{Explore and validate simulated behaviors.} The system necessitates an intuitive presentation of simulation results and supports interactive exploration for users, which ensures that the simulated outcomes are visually accessible and actively engaged by users, fostering a more dynamic and user-driven exploration process.

\item 
\textbf{Analyze social media behaviors further.} To facilitate further research beyond the exploration of behavioral data itself, the system requires additional functionalities to meet more in-depth study requirements. Examples of such requirements include investigating the evolution of social networks, understanding the influence of influential users on the public, and delving into the study of concealed behaviors. These extended capabilities are crucial for conducting comprehensive and nuanced research in the realm of social media dynamics.

\end{enumerate}
\begin{table*}
  \caption{\textcolor{black}{The demographic information of the interviewees. We record their ages, jobs, domains, backgrounds, years of employment, and the frequency of studying social media data.}}
  \label{tab:demogra}
  \resizebox{\textwidth}{!}{
  \begin{tabular}{llllllll}
    \toprule
      ID & Age&Gender&Job&Domain&Background&Years&Frequency\\
    \midrule
     P1 &30& Female&Advertiser&Content creation&Industry&8&Twice a week\\
     P2 &25& Female&Individual media&Content creation&Industry&2&Everyday\\
     P3 &26& Female&Marketer&User behavior research&Industry&4&Everyday\\
     P4 &24& Male&Postgraduate researcher&User behavior research&Academia&2&Everyday\\
     P5 &25& Female&Postgraduate researcher&User behavior research&Academia&3&Everyday\\
    \bottomrule
  \end{tabular}}
\end{table*}

\section{Simulation Workflow}
\begin{figure}[t]
  \centering
  \includegraphics[width=\linewidth]{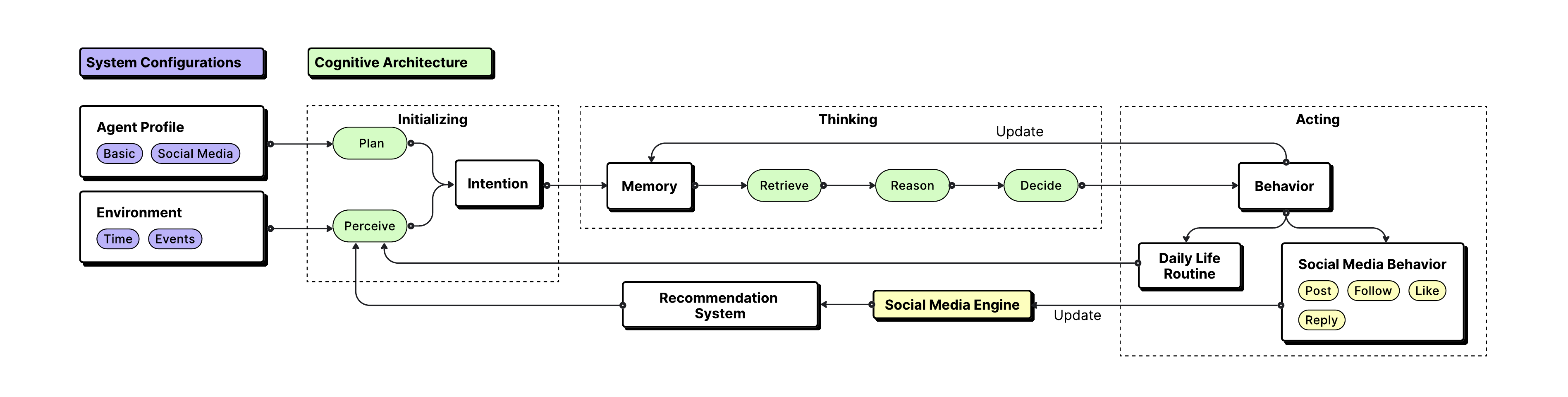}
  \caption{%
    \textbf{The workflow of the simulation model.} It comprises three components: system configuration, social media engine, and cognitive architecture.
  }
  \label{fig:workflow}
\end{figure}
This section introduces the simulation model workflow illustrated in \Cref{fig:workflow}, which was designed to simulate believable social media behaviors~(R1).
To improve believability, we reflected on the prior literature on techniques~\cite{feng_twibot-22_2022,10.1145/3409116} that detect social bots on online platforms that can mimic humans to engage with users, post content, and influence discussions. 
Inspired by the tactics used by those algorithm designers~(feature-based, text-based, and graph-based), we have summarized some noteworthy aspects to promote the believability of generation by focusing particularly on the similarity between agents and real users concerning user features, content style, and social network characteristics.
Feature-based methods~\cite{kudugunta_deep_2018} extract features from users' metadata and apply classification algorithms to distinguish social bots. 
Therefore, we tend to create well-defined characters by setting their demographic features and social habits comprehensively to emulate real social media users.
Text-based methods~\cite{Feng_Tan_Li_Luo_2022} aim to find the difference in language style between genuine users and social bots, so we are committed to generating human-like posts without apparent malice that most social bots are designed for.
Graph-based methods~\cite{10.1145/3308560.3316504} consider social relationships as a graph and utilize graph neural networks to figure out anomalies in social bot networks, which prompted us to pay attention to agents' social interactions, such as "reply", "like", and "follow", with each other. 
Similar to the practice of actors crafting character biographies to inhabit their roles better, we find it necessary to compose individual backgrounds and arrange daily life for each agent meticulously.
Since our behaviors on social media, especially posting, are largely based on our daily lives.
Furthermore, not only should our system generate believable agents, but it should also support the investigation of user behaviors in different scenarios.
To meet the aforementioned requirements, the workflow we proposed comprises three components: system configuration, social media engine, and cognitive architecture.
\re{In this context, the system configuration is set by the user, while social media engine and cognitive architecture are powered by LLMs for automatic operation.
LLMs take text prompts as input, which can range from simple queries to complex paragraphs.
By providing context, LLMs can generate more relevant and coherent responses.
Therefore, by designing specific prompts, LLMs can assist us in performing various computational steps in the simulation process.}

\subsection{System Configuration}\label{config}
Customized by users, system configuration refers to the simulation workflow input tailored to a user's research requirements.
After interviewing some researchers involved in social media studies, we collected and summarized users' requirements for simulation settings. 
We divide system configurations into two parts involving environments and agents.

\subsubsection{Environments}
First, we require users to specify the start and end times of the entire simulation process, as well as the intervals between each behavior simulation.
For example, users have the option to configure the simulation of agents' social media behaviors within a specific week, with simulations occurring at hourly intervals.
Given that users on social media platforms are distributed worldwide, for ease of user configuration, we standardize the adoption of the ``Anywhere on Earth''~(AoE) time zone.
For certain research purposes, users may need to investigate agents' responses to a particular societal public event. 
Therefore, we also support users in specifying the content and timing of public events within the simulated timeframe.
Users can also set some private events that would only be perceived by a specific agent.

\subsubsection{Agent Profile}
To better address research requirements, we provide users with the function to create two types of agents: regular agents and Non-Player Characters~(NPCs).
Regular agents are akin to genuine humans on social media platforms, and users can customize demographic information and social media behaviors for each individual agent.
The demographic information that needs to be configured includes name, age, gender, residency, innate, job, and lifestyle.
Social media habits include followers, posting frequency, primary content of posts, and engagements.
Through the aforementioned configurations, users can flexibly define agents that align with most kinds of research requirements.
Since our simulation workflow is powered by LLMs, users can implement all configurations using text, which provides users with a high degree of freedom.
Additionally, the system can accommodate users in providing either precise or ambiguous descriptions of information thanks to sophisticated language understanding of LLMs.
For example, users have the option to set the age as ``18'', ``from 16 to 24'', or ``adolescents'', or even blank.
NPCs, a concept borrowed from gaming, are utilized in our context to simulate some kinds of individuals or entities with significant influence referred to as key opinion leaders~(KOLs)~\cite{scher2021key} within real-world social platforms~(R4). 
In contrast to regular agents, we do not configure any demographic information or social media habits for NPCs.
Instead, users can establish their specific identity and specify the exact timing and content of posts made by these NPCs.
These NPCs agents can be employed to simulate the deployment of commercial advertisements or the dissemination of public information and guidance on social media.

\subsection{Social Media Engine}
For enhanced simulation, we have developed a virtual social media platform called ``Sparkle'' and designed a recommendation system, since contemporary social media platforms commonly employ recommendation mechanisms to suggest content of interest to users.
``Sparkle'' means refracting light in a way that creates a twinkling effect, suggesting the ignition of creativity, ideas, and interactions within the social platform.
Each post is referred to as a ``spark''.
``Sparkle'' supports functionalities encompassing posting, commenting, liking, and following other agents.
Concerning existing recommendation systems~\cite{10144391,10.1145/3568022}, our designed recommendation system is based on two criteria: recommending posts of interest to users and recommending specific content (such as advertisements posted by NPCs).
More specifically, the posts of NPCs and followed agents will be recommended to the corresponding agents compulsorily~(R4).
For other cases, we utilize LLMs to calculate the strength of the recommendation.
\re{Whenever an agent posts a ``spark'', LLMs will sequentially compute the recommendation level of that ``spark'' for other agents.}
If the output surpasses a threshold \re{set by users}, the post will be recommended to specific agents.
The core of the prompt template is shown below:
\begin{verbatim}
    There is basic information of {Agent1}. {Agent1's Demographic Information}
    {Agent2} posted that {Content} at {Time}.
    On a scale of 1 to 10, where 1 is not recommended and 10 is highly recommended,
    rate the likely that Sparkle recommend the {Agent2}'s post to {Agent1}.
    Rate (return a number between 1 to 10):
\end{verbatim}
The content within “\{\}” can be modified based on specific circumstances.
Additionally, users have the option to customize the recommendation threshold.

\subsection{Cognitive Architecture}
The most crucial component of our simulation workflow is the cognitive architecture, which enables agents to generate corresponding actions based on existing conditions.
We combined a hierarchy of thinking kinds~\cite{rips_folk_1989} and the cognitive architecture proposed by Park\etal~\cite{park2023generative}  with some specific modifications tailored to simulating social media behaviors.
The cognitive architecture is divided into three parts: initializing, thinking, and acting.
Like running a program on the computer, these three parts are considered as input, function, and output, respectively.
During each simulation, we acquire the agent's intention for action, which undergoes a series of thinking processes before a final action is made. 
As mentioned, we are required to generate two types of behaviors: daily life actions and social media behaviors.

\subsubsection{Initializing}
We summarize that the intention for action originated from two aspects, internal planning and external perceiving~\cite{10.1007/BFb0013570}.
Plans describe a forthcoming sequence of actions.
Based on life experience, in the absence of unforeseen circumstances, we are highly likely to act in daily life according to the plan.
\re{Before the agents start their daily routines, we use the LLM to calculate the wake-up time for each agent.}
We utilize LLMs to generate an agent's plan beginning with the wake-up hour once a day by employing the prompt template proposed by Park\etal~\cite{park2023generative}.
The results demonstrate what the agent plans to do at various times during the day.
The other part of intention comes from perception, which exerts a significant influence on social media behavior.
In our model, we categorized perception into three kinds: public events~(specified by users), daily life experiences~(simulated by our workflow), and social media information~(recommended by ``Sparkle'').

\subsubsection{Thinking}
In this part, we maintain a memory system and implement three thinking modules: retrieve, reason, and decide to simulate the thinking process.
It's worth mentioning that we deliberately add a reasoning component to enhance the interpretability of simulated results~(R4).
Reasoning is the cognitive process of systematically analyzing information or evidence to draw logical conclusions, make inferences, and evaluate the validity of statements~\cite{rips_folk_1989}.
We develop the memory system to record the agent’s real-time actions and perception, especially to overcome the short-term memory capabilities of LLMs.
As for the retrieve module, we also adapt the function proposed by Park\etal~\cite{park2023generative}.
When the agent has obtained the intention, it is designed to retrieve relevant memories from the system according to three elements, recency, importance, and relevance. 
Recency is based on memories' temporal proximity to the present moment. 
We implement recency as an exponential decay function since the memory was last retrieved.
Importance measures the significance of memory for the agent, which is calculated by LLMs.
\minor{The prompt template is shown in Appendix~\ref{important}.}
Relevance refers to memories closely connected to the current situation.
We calculate it by the cosine similarity between the memory’s embedding vector and the current situation’s embedding vector.
The retrieval score is considered as a weighted combination of the three scores.
After completing retrieving, the agent will enter the reasoning process to draw logical conclusions and then decide what action it should take.
We leverage LLMs' ability to generate coherent and contextually relevant text to achieve this process with appropriate prompts.
We provide the core of a prompt example of whether the agent decides to post on social media \minor{in Appendix~\ref{decide}}.
Given the prompt, \tocheck{the reasoning and decision will be provided together. } The possible output that the agent would not post will be:
\begin{verbatim}
    {"Reasoning":"Rebecca is a data analyst who loves exploring data trends.
    She usually posts around 1 times a day, and has already posted once today.
    Therefore, it is unlikely that she would post again right now",
    "Answer":"No"}
\end{verbatim}

\subsubsection{Acting}
There are two kinds of acting, daily life behaviors and social media behaviors.
The generation method and prompt templates of daily life behaviors are nearly adapted from Park\etal~\cite{park2023generative}. 
As for social media behavior, we focus on four primary behaviors in real life, ``post'', ``like'', ``follow'', and ``reply''.
\minor{The corresponding prompt core will be presented in Appendix~\ref{acting}.}

\subsection{Workflow}\label{chain}
The simulation workflow requires users to complete system configuration~(\cref{config}) as input.
During each simulation interval configured by users, we first generate the daily life actions of all agents, such as ``having lunch'' and ``going for a hike''.
Then, we will simulate the ``post'' behavior of all agents~(if agents want to post ``spark'' and what they post).
Subsequently, the recommendation system recommends these ``spark''s to all agents. 
The agents then decide if to execute the social media behaviors, including liking, following, and replying, along with the reasons for these actions.
After each simulation, the memory system undergoes real-time updates, with actions output serving as the perception for the next simulation.
It is worth noting that each prompt adheres to ``chain-of-thought'' rules~\cite{NEURIPS2022_9d560961} to enhance the reasoning capabilities and output controllability of LLMs.
\minor{A more concrete description of design rules is provided in Appendix~\ref{achain}.}

\section{System}
\begin{figure}[t]
  \centering
  \includegraphics[width=\linewidth]{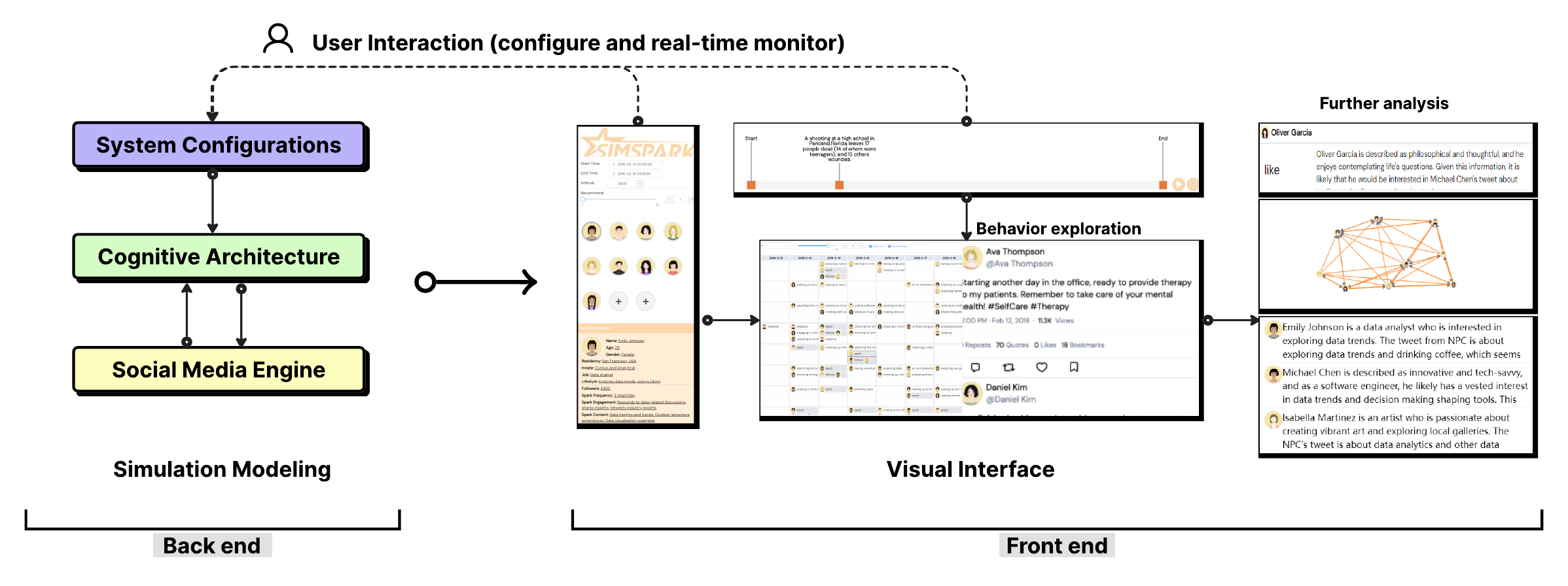}
  \caption{%
    \textbf{The framework of SimSpark} includes simulation workflow~(back end) and visual interface~(front end). The back end simulates social media behaviors. The front end provides visual design and interactions for users.
  }
  \label{fig:framework}
\end{figure}
In this section, we introduce the framework of SimSpark, including simulation workflow~(back end) and visual interface~(front end). 
SimSpark was developed for social media researchers and stakeholders to investigate different behaviors by simulating different scenarios and agents.
We also elaborated visual design and interactions of the interface. 

\subsection{Framework}
The structure of the system framework is shown in \Cref{fig:framework}. 
The ``Visual Interface'' part is a reduced version of \Cref{fig:teaser}, and detailed information can be found in \Cref{fig:teaser}.
Users can configure the system settings and initiate simulations through the front-end interface.
The results of each simulation can be displayed in real-time on the front end for users to monitor the process conveniently.
During the simulation process, if users are dissatisfied with the results, they can pause the simulation at any time and modify the system settings to adjust the simulation outcomes.
For instance, they can add public events or modify agent profiles to align with research requirements better.
The front end also empowers users to analyze generated behaviors deeply.
The simulation model is implemented with Python.
Our implementation utilizes the GPT-4 version.
The front end is written in JavaScript and open-source framework Vue.js.

\subsection{Interface}
We introduce our interface design illustrated in ~\Cref{fig:teaser}, which combines simulation parameters customization and agent behavior analysis.

\subsubsection{Simulation Parameters Customization~(R2)}

 To utilize this system effectively, users are initially required to finalize the parameter settings within Setting Panel~(\Cref{fig:teaser}A). This involves setting the start and end times of the simulation, determining the simulation interval, and establishing the threshold for the ``Sparkle'' recommendation mechanism. For each agent, if a ``spark'' is not posted by NPC or the agents followed, the system employs LLMs to generate a recommendation score for ``spark''s directed toward the agent. If the recommendation score of a ``spark'' for the specific agent falls below the established threshold, that ``spark'' will not be recommended.

 Subsequently, users are required to finalize the settings for agents within~\Cref{fig:teaser}B. 
 \re{As shown in Avatar Panel~(\Cref{fig:teaser}B1), each avatar represents an agent.
 By clicking the avatar, the border of the avatar will be darker and the demographic information of selected agents will be shown in Basic Information Panel~(\Cref{fig:teaser}B2), where users can set the characteristics of agents by inputting natural language.
 }
 By selecting the first “+” icon in Avatar Panel~(\Cref{fig:teaser}B1), a new regular agent can be created. \re{Users can also initial} its attributes within Basic Information Panel~(\Cref{fig:teaser}B2). The avatar can also be changed by simply clicking. These attributes encompass demographic information such as age, gender, name, residency, and job, as well as characteristics including innate traits, lifestyle, and social media information such as the number of ``Sparkle'' followers, ``spark'' frequency, ``Sparkle'' engagement, and ``Sparkle'' content. 
 Furthermore, by selecting the second “+” icon in Avatar Panel, NPCs can be added~(R4). 
 These NPCs can only have their identities, ``spark'' times, and content set within Basic Information Panel.

 Lastly, users are required to finalize the event settings within Timeline~(\Cref{fig:teaser}C). 
 \re{The timeline shows the progress of the simulation.
 Hovering the mouse over the timeline, the specific time represented by the current position will be displayed.
 Users can configure events in this view.}
 \re{First,} by using the multi-select checkbox in the top left corner, users can configure the events to be perceived by either all agents or the specific one.
 \re{Next, }by selecting a specific point on the timeline, a new event can be generated at that moment. Users are then required to input descriptive text for the event.
 \re{The events perceived by all agents are represented by an orange rectangle on the timeline.
 While the events perceived by only one agent are represented by its avatar.
 The text description is shown above the timeline.}
 Upon completion of the event settings, users can initiate the simulation by selecting the play button located in the lower right corner. During the simulation, \re{the timeline displays an animation of the progress bar, and}
 users have the option to pause it by clicking the same button again or restart the simulation by clicking the reset button on the right.

\subsubsection{Agent Behavior Analysis}

 Upon the simulation’s start, all agent behaviors in Calender~(\Cref{fig:teaser}D) are systematically displayed in a schedule~(R3). 
 \re{The horizontal axis of the calendar represents the dates, while the vertical axis represents the hours.}
 These behaviors include daily life activities, which are unhighlighted, and social media behaviors, marked with a gray background. The latter mainly involves ``post'' and ``follow'' actions. The system schedules according to Anywhere on Earth~(AoE) time, leading to apparent time differences in behaviors for agents in different locations. To keep it simple, the system only shows an activity at its start when an agent engages in a continuous activity. For example, the system only shows the sleeping activity when agents just fall asleep, even though it lasts for more than 1 interval.

 In terms of interactivity, users have the option to switch from displaying all agents’ behaviors to the behavior of a specific agent, which can be done via the top left corner of the interface. The importance threshold can be adjusted by sliding the slider located at the top or by modifying the number on its right; behaviors falling below this threshold will be hidden. Furthermore, users can choose to hide behaviors related to daily life or social media activities by checking the box in the top right corner. By hovering over the timeline in~\Cref{fig:teaser}C with their mouse, corresponding cells in the schedule will be highlighted\re{, \ie the border will be orange}, facilitating users to explore the relationship between agents’ behaviors and set events.

 Users are endowed with the capability to interact with specific social media behaviors for more comprehensive details~(R3). 
 The content associated with these behaviors is displayed in Sparkle View~(\Cref{fig:teaser}E). 
 \re{Sparkle View~(\Cref{fig:teaser}E) display the detailed information of the ``spark''. 
 The design of this interface feature draws inspiration from Twitter's design. 
 In the view, users can see the poster, the post content, the posting time, the number of likes and replies, and the agents who liked the post. 
 The detailed reply information~(reply agent and content) is displayed after clicking the dialog box icon at the bottom left.
 All ``spark''s are displayed in chronological order in this view. 
 While, when a social media behavior is clicked in the Calendar~(with a gray background), only the corresponding ``spark'' will be displayed in the Sparkle View to facilitate user exploration.}
 By clicking ``like'' avatar or ``reply'' avatar in Sparkle View~(\Cref{fig:teaser}E) or ``follow'' behavior in  Calender~(\Cref{fig:teaser}D), the rationale behind these behaviors is elucidated in Reasoning View~(\Cref{fig:teaser}F), thereby offering users insight into the motivations that drive the agent’s actions. For instance, the diagram initially outlines the persona of Oliver Garcia. It then infers from the content of the ``spark'' that it aligns with Oliver’s inherent traits and interests, thereby revealing the reason behind Oliver Garcia’s liking of this particular ``spark''.

 Finally, Follow Network~(\Cref{fig:teaser}G) presents the "follow" relationships between agents in the form of a node graph. Users can hover the timeline in ~\Cref{fig:teaser}C to observe how "follow" relationships between agents evolve over time~(R4).
 The view can be switched to Hidden Reason~(\Cref{fig:casexxx}B).
 Click a ``spark'' in Calender~(\Cref{fig:teaser}D), and the details of it will be shown in Sparkle View~(\Cref{fig:teaser}E) and the reasons why agents do not follow, like, reply this post will be shown in Hidden Reason~(R4).
\section{Evaluation}
SimSpark aims to provide a flexible platform to help researchers simulate social media behaviors with customized agents and environmental conditions.
To evaluate the effectiveness of our system, we conducted case studies in different scenarios.
A quantitative study was arranged to illustrate to what extent generated data can approximate natural social behaviors.
We also interviewed some domain experts who perform causality analysis in their research to collect user feedback on our system.

\subsection{Case Study}
We first demonstrated the usability of \textit{\textbf{SimSpark}} through two different cases.

\subsubsection{User Passion and Affect}
\begin{figure}[t]
  \centering
  \includegraphics[width=\linewidth]{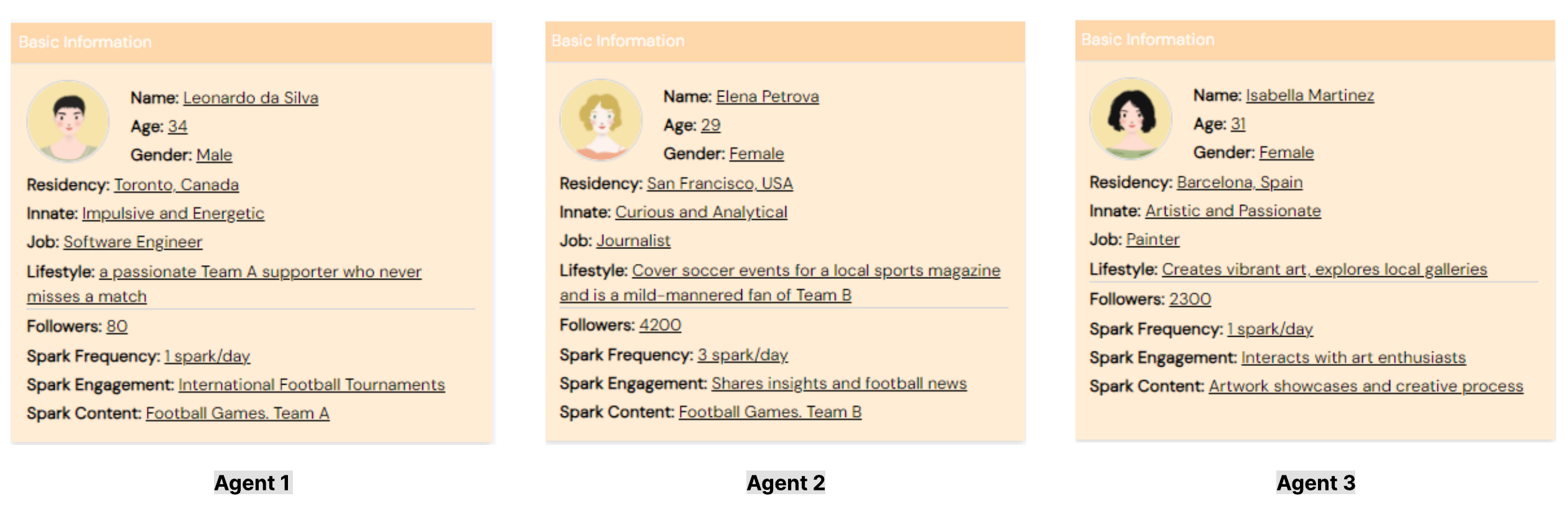}
  \caption{%
    The detailed information of three customized agents.
  }
  \label{fig:agent}
\end{figure}
\begin{figure}[t]
  \centering
  \includegraphics[width=\linewidth]{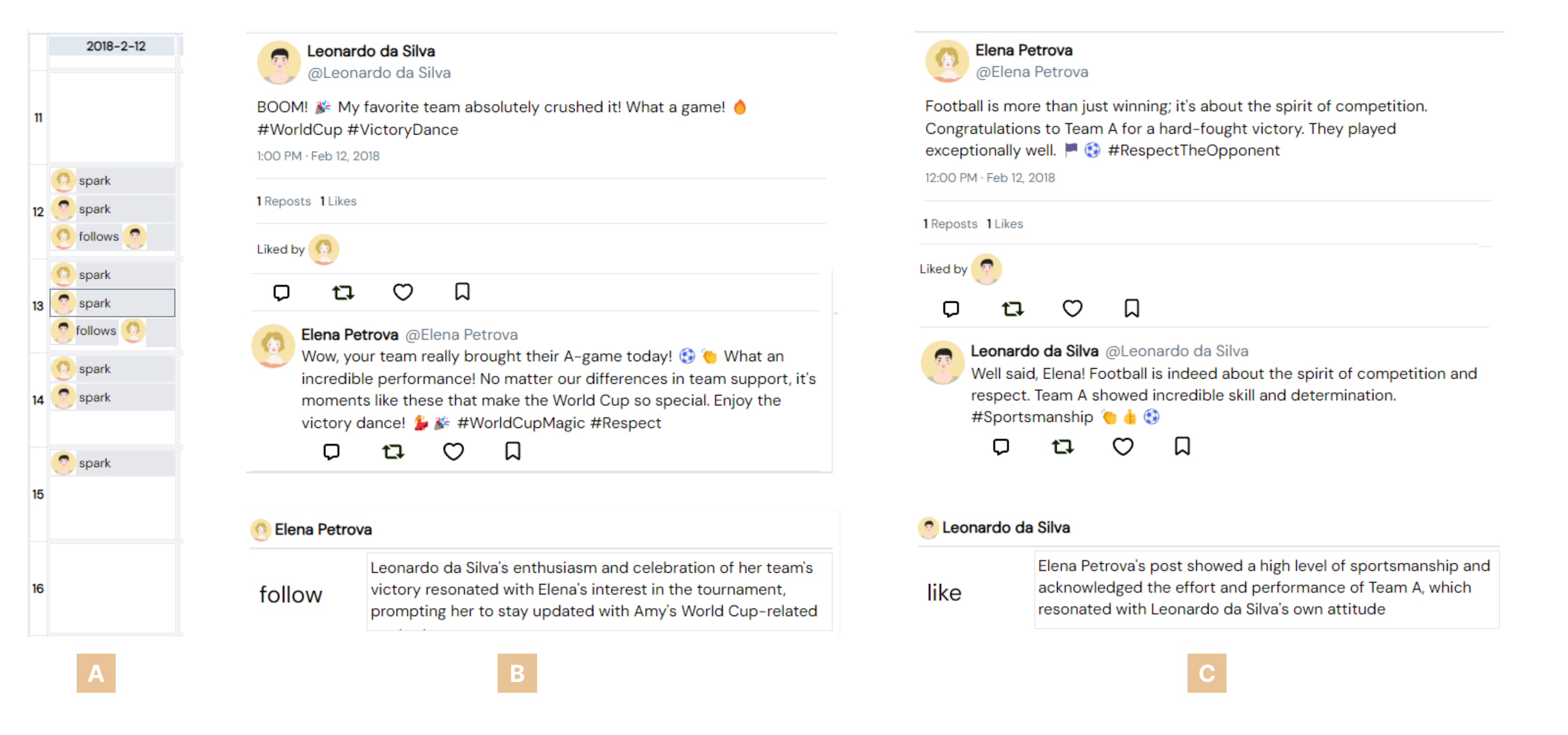}
  \caption{%
    \textbf{A} displays the social media behaviors of agents. \textbf{B} gives an example of Leonardo da Silva's post, Elena Petrova's reply, and the reason why Elena Petrova follows Leonardo da Silva.  \textbf{C} gives an example of Elena Petrova's post, Leonardo da Silva's reply, and the reason why Leonardo da Silva likes this post.
  }
  \label{fig:spark1}
\end{figure}
\begin{figure}[t]
  \centering
  \includegraphics[width=\linewidth]{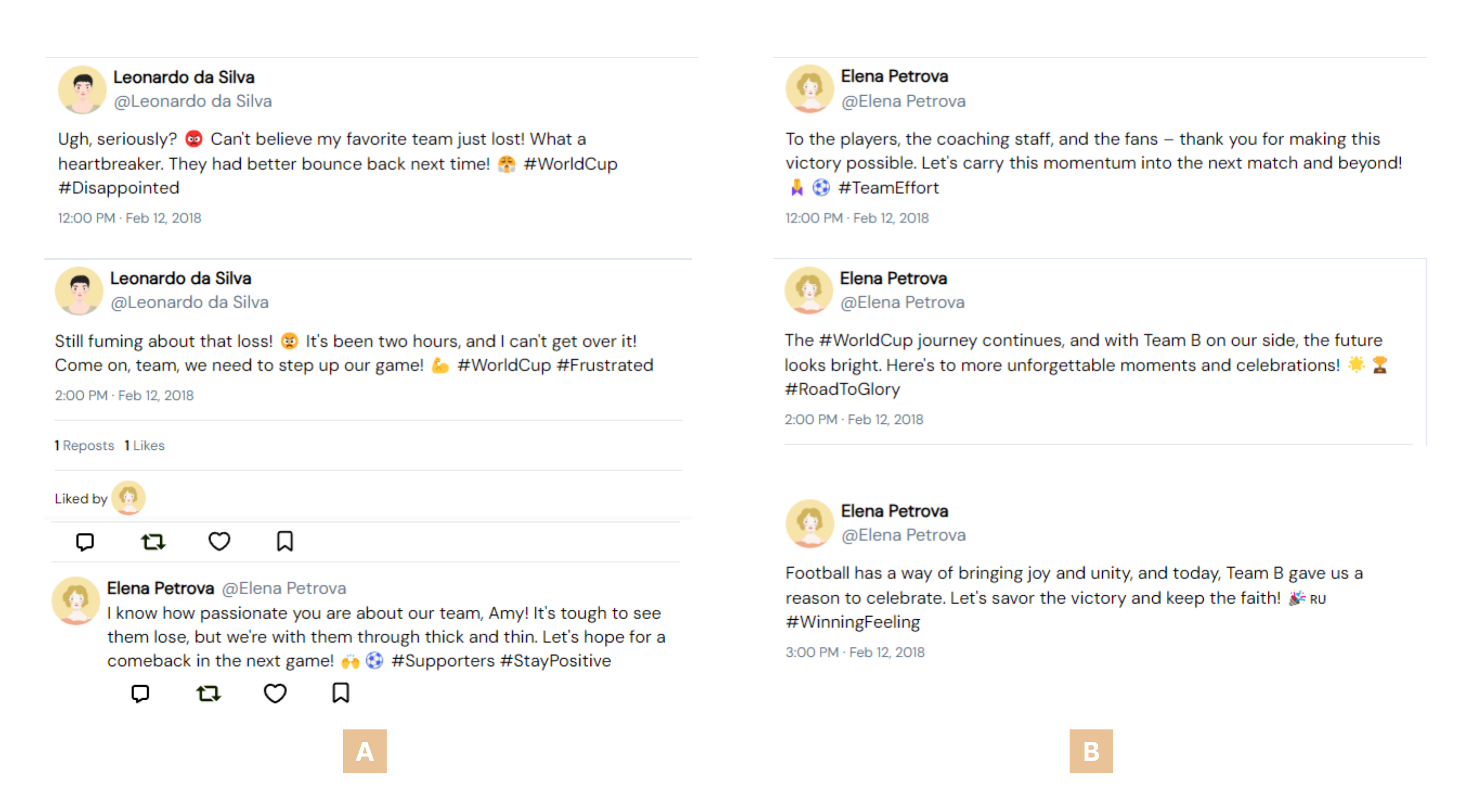}
  \caption{%
    \textbf{A} displays Leonardo da Silva's posts and Elena Petrova's reply. \textbf{B} displays Elena Petrova's posts.
  }
  \label{fig:spark2}
\end{figure}
Football is one of the most popular sports globally.
We created three agents with different attitudes toward football~(shown in \Cref{fig:agent})~(R2).
The first agent, named Leonardo da Silva, is a passionate Team A supporter who never misses a match. 
He is impulsive, energetic and doesn't use social media much.
The second agent, named Elena Petrova, covers soccer events for a local sports magazine and is a mild-mannered fan of Team B.
She often posts on social media, sharing her daily life, and has many followers.
The third agent, named Isabella Martinez, is not a football fan and does not frequently utilize social media.
We employed SimSpark to explore the responses of these three agents to different outcomes of the Team A and Team B matches.
First, we added a public event that "Team A emerged victorious over Team B in the World Cup final, securing the championship." and ran the simulation~(R2).
From Calendar View~(\Cref{fig:spark1}A), we can easily know that Leonardo da Silva and Isabella Martinez both commenced posting content frequently~(R3).
Leonardo da Silva enthusiastically celebrated the victory of Team A, while Elena Petrova posted several ``spark''s expressing disappointment but also filled with sportsmanship and respect.
Due to the similarity of their content, their posts are recommended to each other through our recommendation system.
They reciprocally liked, replied, and followed each other.
In \Cref{fig:spark1}, we display some posts and the reason behind the behaviors.
It is not difficult to observe that, despite their support for different teams, the atmosphere is remarkably harmonious.
In spite of Elena's supported team's loss, she did not vent her anger towards the opposing fans, in accordance with her character profile, a mild-mannered fan.

Instead, we modified the event that Team B won over Team A.
Similarly, both of them commenced posting content regularly.
While there has been a significant change in the content.
Leonardo was so angry about the match result that he couldn't let it go even two hours later, and Elena came to console him, rather than harbor hostility, despite supporting different teams~(\Cref{fig:spark2}A).
Yet, Elena shared her exhilaration, emphasized the team's outstanding performance, and encouraged other fans to celebrate together~(\Cref{fig:spark2}B).
In contrast to the previous situation, neither of them has followed the other.
In both scenarios, Isabella Martinez exhibited almost no social media behaviors, possibly indicative of her lack of interest in football matches.
She just posted once, \textit{"Good night, everyone! Had a productive day painting and exploring beautiful art in Barcelona. Feeling grateful for this artistic journey. \#ArtisticSoul"}, since she likes to create vibrant art and explore local galleries.
Wakefield\etal~\cite{wakefield_social_2016} stated that passion for an event positively influences content creation and sharing in social media.
To a certain extent, our case obtained qualitatively similar results to their experiments. 
Isabella Martinez lacks enthusiasm for football; therefore, she exhibits no discernible reaction to whichever team wins.
Leonardo da Silva frequently posts and interacts with other users after the match, whereas his regular posting frequency is quite low.

\subsubsection{Product Promotion}
\begin{figure}[t]
  \centering
  \includegraphics[width=\linewidth]{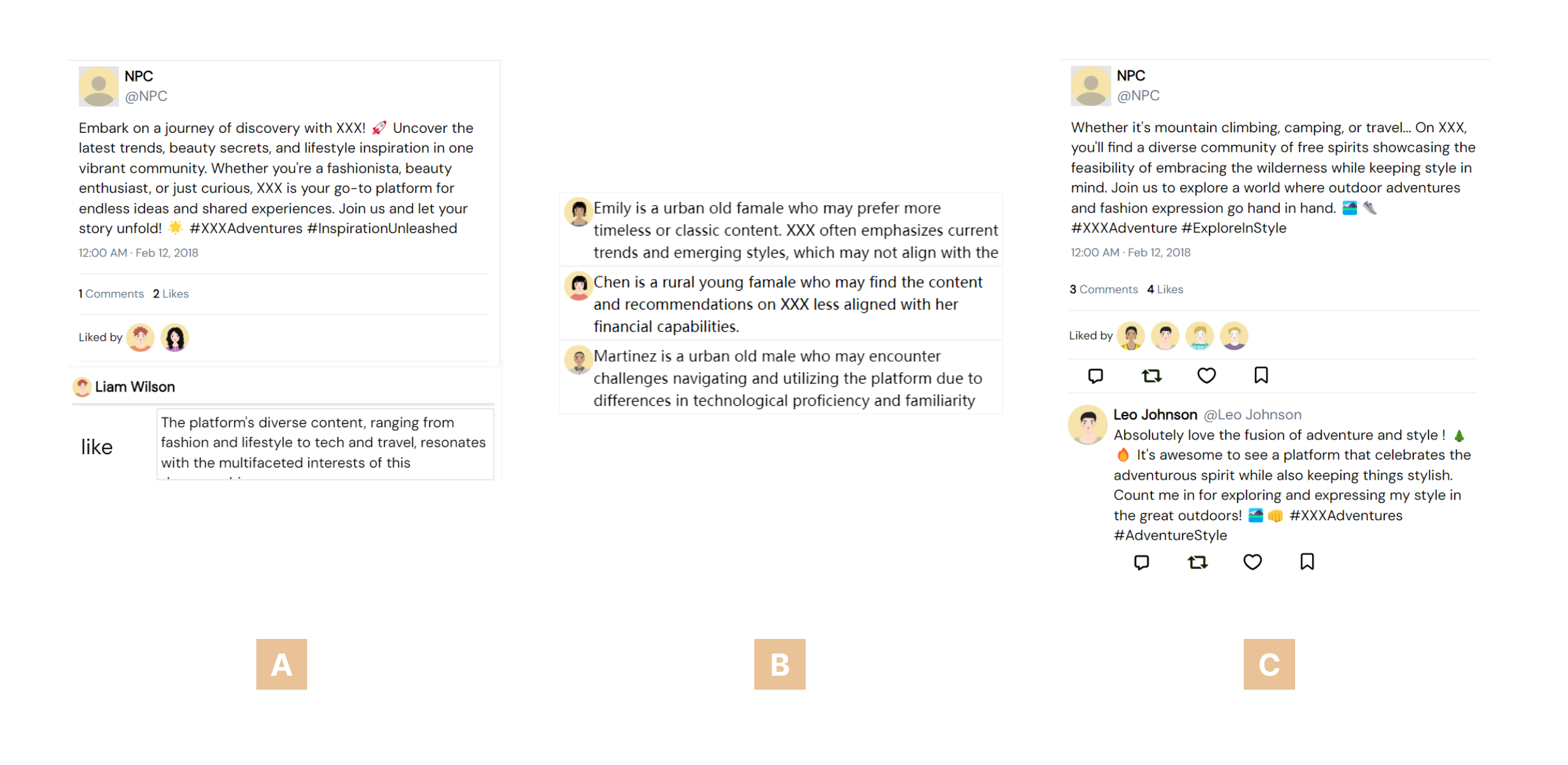}
  \caption{%
    \textbf{A} shows an example of advertising and the reason why the urban young male ``like'' it. \textbf{B} displays part of the reasons why some agents don't show interest. \textbf{C} displays targeted advertising and agents' ``like'' and ``reply'' behaviors.
  }
  \label{fig:casexxx}
\end{figure}
Advertising in social media usually involves the utilization of data analytics, market segmentation, and behavioral targeting to enhance the precision and efficacy of promotional efforts. 
Our system can allow relevant stakeholders for a comprehensive assessment of the effectiveness of advertising by mimicking the engagement patterns and reactions of the targeted demographic~(R2, R4).
We collaborated with an industry professional to jointly complete this case study.
Assume that you are promoting a social app ``XXX'' product where users can discover and share product recommendations, travel experiences, beauty tips, and lifestyle content. 
It is primarily used by young females in first-tier cities, and the objective is to attract users with different characteristics to augment the user base.
We segmented users into eight groups based on city tier, age, and gender, with city tier divided into urban and rural, age categorized into young and senior, and gender distinguished between male and female. These three features were combined to create the segmentation. 
We configured 8 distinct agents with these three features and designed an advertisement for ``XXX''.
We found out that urban young males and urban young females show great interest in this app, while others do not~(\Cref{fig:casexxx}A).
The reasons for non-engagement include a lack of interest in fashion trends, mismatched spending capacity, and difficulties in operation~(\Cref{fig:casexxx}B).
Next, we narrowed down the potential target users to urban young males and devised a more targeted promotion.
We created 6 urban young male agents with different jobs, innate, and lifestyles and advertisements were tailored for them, incorporating elements such as sports, camping, and other related activities.
The positive engagement of the post by agents served as a partial validation of the effectiveness of our marketing strategy~(\Cref{fig:casexxx}C).

\subsection{User Study}
We conducted a quantitative study to validate the realism of our simulation model outputs by testing whether participants can distinguish agents from real humans based on their social media behaviors~(R1). \minor{Our user study adhered to ethical standards, ensuring informed consent and data privacy while paying special attention to participants' autonomy and emotional well-being.}

\subsubsection{Study Procedure}
We designed an experiment to test whether participants could distinguish between generated agents and genuine humans.
We created 20 different agents' profiles randomly by \re{LLMs}.
\minor{The design strategy of prompts can be found in Appendix~\ref{auser}.}
\re{The generated agents} belong to different age groups, have various professions, and follow distinct lifestyles and social media habits.
We generated 7-day routines for all agents, including daily life and social media information, by our workflow.
Next, we recruited 20 genuine social media users from different backgrounds who consented to provide their social media information to us for study. 
\minor{We provided all users with a comprehensive explanation of the purpose, procedures, and potential impacts. In addition, we informed them that they could withdraw from the study at any point, especially if they experienced discomfort during the process.}
\re{Each user was paid \textdollar 10.}
We selected the social media behaviors of both generated agents and genuine humans over two specific days, including posting, liking, replying, and following. 
Additionally, we conducted \re{structured} interviews with those genuine users to understand the reasons behind these behaviors, like we obtained the reasoning process of agents.
For each social media behavior of genuine users, we asked them why they engaged in that behavior, such as \textit{``Why did you post certain content?''} or \textit{``Why did you make the reply?''}
This way, we can obtain the behaviors of genuine users and the reasons behind them.
\minor{We provided one-to-one correspondence of actions and reasons of humans and agents to participants for further study.}
We aimed to determine whether these behaviors and the underlying reasons behind them over these two days, can be discerned as originating from real users by participants.
Furthermore, agents without simulating daily life and without setting social habits are also generated to be applied for the same experiment, which can validate the efficacy of our full simulation workflow. 
The four groups of subjects used in our experiment are as follows:
\begin{itemize}
    \item \textit{Real:} We recruited 20 genuine social media users, with 10 of them identifying themselves as male and 10 as female. 
    Their ages range from 18 to 28 years old, with an average age of 22.3, and they exhibit a high level of proficiency in using social media.
    Because our simulation generates text-only content, we tend to select users who are inclined to publish original textual content and disregard image-based content.
    \item \textit{Agents:} We used the full simulation workflow to generate social media behaviors of the 20 randomly created agents.
    \item \textit{Agents without daily life:} We only simulated social media behaviors of the same agents without generating daily life routines, which implies that their perception is lacking a portion.
    \item \textit{Agents without social habits:} We simulated agents' behaviors without setting their social media habits, involving post frequency, post content, and engagement.
\end{itemize}
We conducted three sets of classification experiments,  
\textit{Real} and \textit{Agents}, \textit{Real} and \textit{Agents without daily life}, \textit{Real} and \textit{Agents without social habits}.
For each experiment, we recruited participants to distinguish 10 subjects sampled randomly from all subjects in the corresponding group, with 5 of them real and the rest being agents. 
To ensure minimal interference in the experiment and protect user privacy, we have anonymized all sensitive information such as names, times, and locations.
\minor{We implemented strict data anonymization and de-identification processes to prevent any disclosure or association of personal information.}
Participants assessed whether the experimental subjects were real users or agents based on their behaviors over two days and were encouraged to provide reasons for their judgments.

\subsubsection{Participants}
We recruited 25 participants from the university community for each experiment, a total of 75 participants, whose ages ranged from 18 to 25 years old, with an average age of 23.6. 
31 of them identified themselves as female and 44 as male.
\re{The participants provided consent by agreeing to a consent form that indicated they agreed to participate in the experiment and allowed the results to be used for subsequent research.}
They were required to be fluent in English and \re{use social media for more than half an hour each day}.
They were also well-educated, with 80 percent of them holding a bachelor's degree and the rest a high school diploma.
\re{It is important to note that they do not have to be our target users.}
They spent approximately one hour per day using social media, thereby gaining a fundamental understanding of the characteristics of social media and encountering posts made by social bots.
As a result, they had a clear understanding of the purpose of our experiment and exhibited a high level of discernment.
\re{We determined the number of participants based on the pilot study and ran a power analysis with $\alpha = 0.05$ and $power = 0.8$.}
They were paid at the rate of  \textdollar10.00 per hour.

\subsubsection{Result}
Our experiments can be considered as a binary classification task.
For the first experiment, participants were required to distinguish regular agents from real users. 
The average error rate was 43.34$\%$, slightly less than 50$\%$, which can illustrate the difficulty in distinguishing the agents we have generated to a certain extent.
The concrete results are shown in \Cref{tab:commands}.
We can indicate that our full simulation workflow outperformed the other two ablated models.
\begin{table*}
  \caption{Error Rate Results}
  \label{tab:commands}
  \begin{tabular}{ccl}
    \toprule
     & Description & Error Rate\\
    \midrule
    1 & Real \textit{vs.} Agents&M=43.34$\%$; SD=5.4$\%$ \\
    2& Real \textit{vs.} Agents without daily life& M=31.66$\%$; SD=3.6$\%$\\
    3&  Real \textit{vs.} Agents without social habits& M=32.76$\%$; SD=4.6$\%$\\
    \bottomrule
  \end{tabular}
\end{table*}
In addition, we conducted a one-way ANOVA test to quantify the difference in the performance of participants among three experiments~(\Cref{tab:test}). 
\re{The first experiment is \textit{Real vs. Agents}, 
the second experiment is \textit{Real vs. Agents without daily life}, and the third experiment is \textit{Real vs. Agents without social habits}.
We compared the results of the three experiments pairwise and conducted a one-way ANOVA test respectively.
For each comparison, the independent variable has two conditions: Experiment 1 and Experiment 2, Experiment 1 and Experiment 3, Experiment 2 and Experiment 3.
The dependent variable is error rates.}
\begin{table*}
  \caption{ANOVA Test Results}
  \label{tab:test}
  \begin{tabular}{ccl}
    \toprule
      Description & F&p\\
    \midrule
     Experiment 1 \textit{vs.} Experiment 2&3.25& 0.034\\
    Experiment 1 \textit{vs.} Experiment 3& 1.68& 0.048\\
      Experiment 2 \textit{vs.} Experiment 3& 0.52& 0.47\\
    \bottomrule
  \end{tabular}
\end{table*}
ANOVA test also confirmed that the error rate of the first experiment was significantly \re{higher} than the other experiments, which demonstrated outstanding performance of our full simulation architecture~(\((p<0.05)\)).
In addition, we can indicate that there is no apparent distinction between the error rates of experiment 2 and experiment 3.

\subsubsection{Analysis} 
According to the results, participants in the first experiment encountered challenges when attempting to distinguish between real users and virtual agents. This difficulty arose due to the high level of consistency observed in the content of posts generated by agents concerning their designated identities, as well as the striking similarity in their social habits, including emotional expression, posting frequency, and tone of communication when compared to real users. 
Notably, some participants noted that emojis in the posts of certain agents contributed to their deceptive nature. Some agents would deliberately remove the subject and add the corresponding tag in their posts to be closer to the posting habits of real users. 
In contrast, participants in the second and third experiments were more likely to distinguish virtual agents from real users. 
In the second experimental group, due to the lack of daily life, there is a notable incongruity between the content of agents' posts and their purported career settings, and the emotional changes between posts span a wide range of emotions, which makes it easy for participants to distinguish them. 
For instance, after posting a heartfelt prayer in response to a tragic shooting incident, one of these virtual agents followed up with posts containing excessively positive statements such as "Great news!" and "Happy coding everyone!" 
In the third group of virtual agents, which were intentionally simulated without social media habits settings, there are extreme cases of post frequency, surpassing 10 posts in a single day. 
Additionally, they engaged in numerous "likes", and some of their posts exhibited repetitive or even identical content. Such aberrant social behavior, distinct from typical user patterns, enabled participants to recognize them as virtual agents readily.
The results of Experiments 2 and 3 also indicate that if existing frameworks~\cite{10.1145/3526113.3545616,park2023generative} are used to simulate social media behaviors without targeted modifications, the simulation outcomes often fall short of expectations.
Since we first proposed simulating daily life actions before social media behaviors and configuring social habits for agents.
In conclusion, the findings from the ablation experiments underscore the pivotal and constructive role played by the simulation of daily life and the specification of social habits within the full simulation workflow.

\subsection{Expert Interview}
Our system is designed for domain experts or stakeholders to simulate social media behaviors in different scenarios~(R2, R3, R4).
To further evaluate our system, we collected qualitative feedback from \re{four} expert interviewees who had a need for the study of social media behavior and had no prior knowledge of our systems. 
\subsubsection{Experts.}
\minor{We required experts to be in our target user group and have extensive experience researching social media data.}
We searched for \minor{qualified experts} through the Internet and recruited them by sending invitations through social media.
One is a researcher studying the dissemination of information on the internet\re{~(E1)}, another is a professional engaged in promoting products on social media\re{~(E2)}, the third is a practitioner in the media industry\re{~(E3)}, \re{and the fourth is a postgraduate researcher who studies rumor spreading~(E4).}
\re{Their detailed demographic information is displayed in Table \ref{tab:demogra2}.}
\re{\subsubsection{Procedure.} The study was composed of three sessions lasting around one hour and a half. We}
first introduced our system and presented cases mentioned in the former section, \re{invited them to use the system}, and conducted a semi-structured interview at the end. 
\re{In the interviews, our questions primarily focused on the following aspects.
First, we wondered about the overall evaluation of our system and if it would be helpful for their work.
Second, we focused on their evaluations of the simulation workflow and system design.
For example, we asked the question \textit{``How do you like our simulation workflow?''}
As for the system, we asked them questions like \textit{``Do you find our system easy to learn? If not, which specific part do you find challenging?''}, \textit{``How do you evaluate our visual design?''}, \textit{``Is the interaction method reasonable?''} and so on.
Finally, we asked them about any other strengths and weaknesses of our work and areas that need improvement.
\minor{As the interviews progressed, we noticed a gradual decrease in new insights obtained. 
By the time we interviewed the fourth expert, most perspectives had already been mentioned by previous interviewees.}
After the interview, they were paid \textdollar 50 per hour.
Subsequently, we collected and analyzed their feedback.
We applied the emergent coding approach based on grounded theory~\cite{noauthor_qualitative_2019}.
The coding method consists of three stages: open coding, concept development, and grouping concepts into categories.
In the first stage, we read through the text, trying to identify the opinions and find a term to describe them, such as ``impressive simulation method'' and ``intuitive calendar design''.
In the concept development stage, collections of codes that describe similar contents are grouped together to form higher-level ``concepts'', such as ``effectiveness of visual design''.
Finally, we summarized the results into four categories: workflow, visual design and interaction, usefulness and useability, and improvement.}

\re{\subsubsection{Qualitative Results.}
After the interviews, we collected their feedback and summarized it as follows:
}

\textbf{Workflow} 
Our interviewees highly appreciated our creative simulation workflow and were impressed by the formidable capabilities of the LLMs. 
They commended the realism of the simulated agents and the authenticity of their behaviors. 
The cognitive architecture of agents is closely aligned with genius users.
\textit{``We were truly impressed by the innovative approach of your simulation workflow; it offers a promising avenue for delving deep into the intricacies of online social dynamics.''}
They expressed enthusiasm about the potential for future research using this simulation workflow.

\textbf{Visual Design and Interaction} 
During the interview, experts praised the effectiveness and expressiveness of visual encoding, especially the calendar design. 
One of the interviewees particularly stated \textit{``I have a strong preference for the design of our avatars.''}
As for interactions, interviewees indicated that the interactions are straightforward, and the learning curve is rapid.
In addition, interviewees appreciated the ability to customize demographic information and social media behaviors of agents, which allowed them to tailor the simulation to specific research questions and scenarios, providing flexibility in study design.
Text input afforded a high degree of freedom. \textit{``I can express myself and convey information without constraints, allowing for a versatile and unbounded means of communication. ''}
The real-time display of simulation results on the front end was also highly valued.
Interviewees mentioned that this functionality allowed them to monitor and assess the simulation's progress and outcomes, enabling them to make real-time adjustments.

\textbf{Usefulness and Usability} 
Interviewees were able to understand and use the system quickly after the brief introduction.
Experts affirm that our work will have great promise in a variety of fields.
\textit{``It exceeds my expectations in terms of its functionality and contributions to its objectives.''}
They thought highly of the ease with which users can interact with and navigate the system. 
The intuitive and user-friendly system facilitated smooth and efficient user experiences.
\textit{``Its user-friendly interface made it a pleasure to work with.''}
Additionally, interviewees reflected that the method proposed by us offers a novel approach to studying social media behavior. 
\re{Rumor researcher said \textit{``This system allows me to easily test the effects of different intervention methods on the spread of rumors.''}}
The professional engaged in promoting products said, \textit{``We typically conduct research on user preferences through questionnaires.''}
The drawbacks of questionnaires include potential response biases, limited depth of qualitative insights, issues of sample representativeness, as well as the potential for respondent fatigue, which may compromise the reliability of collected data.
\textit{``SimSpark can, to a certain extent, mitigate these issues.''}
Our model can customize environments, and the results can accurately reflect users' genuine needs, thereby circumventing the biases associated with traditional survey questionnaires.
Furthermore, the interviewees appreciated our attention to the reasons behind agents refraining from certain behaviors, which held research value in specific studies as well.
\textit{``We can draw insights for improving marketing strategies from the reasons why agents refuse to like or follow us.''}
\re{One of the interviewees posited that our system not only displays the reasons for ``decide-to-do'', but also the reasons for ``decide-not-to-do'', which holds greater
research value.
\textit{``By analyzing `decide-not-to-do' reasons, I can understand the reasons why users do not take action against rumors, which can help in developing methods to prevent the spread of rumors.''}}

\textbf{Improvement}
Our interviewees also provided us with some meaningful suggestions. 
\textit{"Customizing individual agents one by one can be cumbersome; therefore, there is a desire for automated generation methods and the function for us to make fine-tuned adjustments."}
Additionally, another interviewee suggested that we could generate some public events automatically according to the simulation time set by users.  
\re{Futhermore, some interviewees pointed out that the agents currently generated by the system were not yet scalable. \textit{``If the system can generate agents in large quantities, I can use it to test more diverse scenarios.''}}
 
\begin{table*}[h]
  \caption{\textcolor{black}{The demographic information of the experts. We record their ages, jobs, domains, backgrounds, years of employment, and the frequency of studying social media data.}}
  \label{tab:demogra2}
  \resizebox{\textwidth}{!}{
  \begin{tabular}{llllllll}
    \toprule
      ID & Age&Gender&Job&Domain&Background& Years&Frequency\\
    \midrule
     E1 &27& Female&Research&User behavior research&Academia&4&Everyday\\
     E2 &25& Female&Advertiser&Content creation&Industry&2&Everyday\\
     E3 &24& Male&Media practitioner&Content creation&Industry&2&Everyday\\
     E4 &24& Female&Postgraduate researcher&User behavior research&Academia&3&Twice a week\\
    \bottomrule
  \end{tabular}}
\end{table*}
\vspace{-0.5em} 
\section{Discussion}
In this section, we conduct a discussion on \re{the implication, the boundaries of agents' behaviors,} the application of our system,
the ethics issues raised by simulation, the limitations, and the promising future directions of our work.

{\subsection{Implications}
In this paper, we address the research question of how human-computer interaction can support more flexible and customizable social media behavior simulations to better meet the research needs of target users.
First, We customize the simulation workflow for social media behaviors, leveraging cognitive models to extend the capabilities of the LLM.
Second, we design a user-friendly interface to facilitate user control over the simulation, involving implementing simulation configurations and a real-time monitoring process.
It allows users to test different simulation scenarios and see their effects in real-time.
Next, we provide a set of visualizations to display simulation results to make them easily understood.
It also supports users in exploring the simulated data and behaviors dynamically.
We evaluate our system through case studies, a quantitative experiment, and expert interviews. 
Case studies indicate the potential usage scenarios of our system.
Our quantitative experiment evaluates the believability of behaviors simulated by our workflow.
Expert interviews evaluate the effectiveness and efficiency of our overall system by our target users.
\minor{Notably, our work emphasizes the diversity of social media behaviors and the interactivity of the simulation platform.
Our approach explores a broader range of user behaviors beyond mere posting actions~\cite{10.1145/3526113.3545616}, which captures the nuanced ways individuals engage on social media.
Furthermore, our platform is designed to be highly interactive, allowing users to manipulate simulation parameters and test various scenarios directly, compared to existing simulation systems~\cite{gao2023s3,lin2023agentsimsopensourcesandboxlarge}. }

\re{\subsection{Boundaries}
Although our simulation workflow strives to replicate the decision-making processes of human social media behavior as accurately as possible, certain simplifications have been made. 
Consequently, the agents' behaviors have some limitations that future work can focus on addressing.
We outline the boundaries of our agents compared to humans in the following three aspects.
\subsubsection{Thinking Processes}
We simulate the thinking process by maintaining a memory system and implementing three thinking modules: retrieve, reason, and decide.
Although we leverage the powerful generative capabilities of LLMs to enhance these components beyond the performance of simple rule-based methods, they might not capture complex decision-making processes.
For instance, we have overlooked the reflection module in the thinking processes.
A reflection module allows humans to evaluate their past actions, learn from their experiences, and adjust their future behaviors accordingly.
Our process merely incorporates the agents' past behaviors into their memory system without deeply summarizing lessons learned, which prevents the agents from evolving.
Since we are only simulating the behavior of agents for a few days, the absence of this process is temporarily acceptable.
In future work, incorporating this module can be considered to simulate human thinking processes better, which can significantly enhance the realism, flexibility, and effectiveness of simulations.
\subsubsection{Language Habits.}
One of the primary outputs of our model is the content posted by the agents.
As shown in the cases we provided~(\Cref{fig:casexxx,fig:spark1,fig:spark2}), our generated content includes hashtags and emojis, aligning with a common expression habit on social media platforms.
However, we find that the contents we generated have relatively complete structures and a tone that is more formal and polite, which differs somewhat from the language typically used on real social media platforms.
Real users develop the habit of summarizing their thoughts and using abbreviations, acronyms, and hashtags to convey messages efficiently in today's era of rapid information dissemination.
For example, users frequently use acronyms~(\eg  ``LOL'' for ``laugh out loud'') and abbreviations~(\eg ``BTW'' for ``by the way'') to save space and time.
Furthermore, users often use an informal tone and slang, creating a casual and approachable atmosphere.
Memes and internet culture have become a significant part of online communication.
To adopt such language habits, future work could involve training specialized language models to better align with real human behavior on social media.
\subsubsection{Social Network.}
In this research, we focus on the dynamic ``following'' network among agents.
They can stay informed about current events, trends, and topics of interest from agents they follow.
Information is disseminated through such a network.
However, social networks are much more complex in real life.
The network's structure can also be influenced by factors such as clustering, community formation, and the presence of influencers and hubs.
These factors shape how information flows, how users interact, and how communities form and evolve.
Communities reflect shared interests, common activities, or similar characteristics among users.
Influencers can shape public opinion and consumer behavior, making them valuable for marketing and communication strategies.
The analysis of community detection and key nodes is crucial for social network research; however, we did not conduct a more in-depth analysis in this area, which can be a promising future direction.
}

\subsection{Applications}
The application of simulating social media behavior spans a wide range of fields and objectives, leveraging the capabilities of simulations to gain insights and achieve specific aims. We illustrate it with two examples as follows:

\textbf{Marketing and Advertising} Businesses use simulations to predict consumer behavior and assess the effectiveness of marketing campaigns on social media platforms. 
This enables companies to optimize their advertising strategies and target specific customer segments.
For example, if a retail e-commerce company wants to optimize its marketing strategy for an upcoming holiday season. 
Using past demographic information, the company can create a simulation model replicating consumer behavior during holiday seasons.
Before launching their holiday marketing campaigns on social media, they can simulate the impact of different advertising creatives and campaign strategies on consumer engagement and conversion rates.
Moreover, the company can forecast potential sales outcomes for the upcoming holiday season by referring to consumers' simulated reflections on social media, which enables them to estimate how different marketing strategies will affect sales, revenue, and return on investment (ROI).
In summary, the e-commerce company can make informed decisions and optimize its marketing strategies to maximize its holiday season sales and profitability by utilizing simulations to predict consumers' behavior.

\textbf{Policy and Development} Governments and policymakers can utilize simulations to analyze the potential outcomes of policies for more effective decisions. When they have several policy options to consider, they can use our system to simulate public attitudes corresponding to the option by taking into account the demographics of the population. 
For example, a government health department is faced with the challenge of reducing the spread of a contagious disease within its population.
They run simulations for different scenarios, each representing a combination of policy interventions.
One scenario might involve implementing strict quarantine measures and widespread testing, while another may focus on vaccination and public awareness campaigns.
Policymakers use the simulation results to compare the public's reactions to different policy combinations and anticipate shifts in public opinion, which Facilitates the implementation of targeted public opinion control measures.

The applications of simulating social media behavior are extensive and span various industries and disciplines, which provides a valuable tool for understanding, predicting, and influencing human behaviors and interactions in the dynamic and interconnected world of social media.

\subsection{Ethics Issues}
Simulating social media behaviors will raise important ethical risks.
The generated agents can inadvertently perpetuate biases and stereotypes present in the training data. 
LLMs are trained on the vast corpus from the Internet, which may contain inherent biases. 
These biases can include stereotypes related to race, gender, ethnicity, and other characteristics. 
When the model is trained on such data, it can inadvertently incorporate these biases into its understanding of social media behaviors.
If the training data contains biased behaviors, the simulated agents may replicate these biases in their interactions, thereby perpetuating stereotypes and even unintentionally amplifying them.
Furthermore, if not properly controlled and designed, agents can engage in actions that are discriminatory or offensive, mirroring the biases in the data.
Therefore, ethical considerations place the responsibility on researchers and developers to actively mitigate and counteract these biases and avoid reinforcing discriminatory or harmful behaviors. 
This may encompass preprocessing the training data to eliminate biases, formulating algorithms that mitigate biases, and establishing protective measures to prevent the perpetuation of discriminatory conduct.
Promoting algorithmic fairness ensures that simulated agents do not contribute to the marginalization or harm of specific groups.
However, it is challenging to test biases and stereotypes comprehensively, and further exploration by researchers is warranted.

In addition, the utilization of social behavior simulations requires strict regulation. 
If these agents are indiscriminately deployed on social media platforms, they can generate content that resembles misinformation or disinformation.
Spreading propaganda or influencing public opinion can have serious ethical implications, which infringe upon individuals' autonomy and their right to make informed decisions based on accurate and unbiased information. 
What's worse, manipulative use can exacerbate social polarization by amplifying extreme views or fostering divisiveness, further fragmenting online communities.
If the agents interact with real users, there is potential for deception.
Those agents are capable of mimicking genuine users for harmful purposes to erode trust in online communication platforms.

\subsection{Limitations and Future Work}
We have comprehended several limitations of our system and suggest possible solutions for them. 
First, we have simplified the rules of real social media platforms in implementation.
We only support four kinds of behaviors and text content output.
However, visual content, including images and videos, has become central to contemporary social media, which is capable of supporting more complex operations.
We have additionally streamlined the guidelines of the recommendation system.
For further study, the social media engine requires a more sophisticated design and supports more user interactions to simulate real-world platforms better.
Second, our system is designed with a focus on agents at its core for deep behavior analysis, yet it overlooks the information propagation of public social platforms. 
Specifically, when researchers need to conduct studies on online public opinion and information diffusion, text or network visualization design can be considered to apply for supporting the analysis of relevant issues. 
Additionally, the simulation model proposed in our study exhibits potential for refinement in improving its stability and versatility due to the frequent utilization of large language models.
Integrating parallel computing and advanced algorithms could be explored to enhance the efficiency and scalability of the system.
\minor{Lastly, we generated 20 agents to evaluate the believability of our simulation framework in the user study. Due to the limited scale of our experiment, we chose not to generate additional agents. This may introduce some degree of randomness into our experimental results. We plan to expand the experiment scale to further demonstrate the robustness of our approach in future.}

\section{Conclusion}

We propose ${SimSpark}$, an interactive system that simulates social media platforms, enabling the study of social media user behaviors in a controlled, ethical manner. ${SimSpark}$ addresses key challenges in generating believable behaviors, validating simulation results, and providing interactive control for generation. It employs a simulation workflow leveraging LLMs to create realistic agent behaviors. The system also offers visualizations for data analysis and supports real-time parameter adjustments for customized settings. Its effectiveness has been demonstrated through case studies, quantitative assessments, and expert interviews. However, our current system still has some limitations, such as the simplification of real social media platform rules. In future work, we would like to further improve ${SimSpark}$ by handling the above limitations.

\begin{acks}
We want to thank the reviewers for their suggestions. This work is supported by Natural Science Foundation of China~(NSFC No.62472099 and No.62202105).
\end{acks}
\bibliographystyle{ACM-Reference-Format}
\bibliography{sample-base}


\begin{thebibliography}{75}


\ifx \showCODEN    \undefined \def \showCODEN     #1{\unskip}     \fi
\ifx \showISBNx    \undefined \def \showISBNx     #1{\unskip}     \fi
\ifx \showISBNxiii \undefined \def \showISBNxiii  #1{\unskip}     \fi
\ifx \showISSN     \undefined \def \showISSN      #1{\unskip}     \fi
\ifx \showLCCN     \undefined \def \showLCCN      #1{\unskip}     \fi
\ifx \shownote     \undefined \def \shownote      #1{#1}          \fi
\ifx \showarticletitle \undefined \def \showarticletitle #1{#1}   \fi
\ifx \showURL      \undefined \def \showURL       {\relax}        \fi
\providecommand\bibfield[2]{#2}
\providecommand\bibinfo[2]{#2}
\providecommand\natexlab[1]{#1}
\providecommand\showeprint[2][]{arXiv:#2}

\bibitem[Ali~Alhosseini et~al\mbox{.}(2019)]%
        {10.1145/3308560.3316504}
\bibfield{author}{\bibinfo{person}{Seyed Ali~Alhosseini}, \bibinfo{person}{Raad Bin~Tareaf}, \bibinfo{person}{Pejman Najafi}, {and} \bibinfo{person}{Christoph Meinel}.} \bibinfo{year}{2019}\natexlab{}.
\newblock \showarticletitle{Detect Me If You Can: Spam Bot Detection Using Inductive Representation Learning}. In \bibinfo{booktitle}{\emph{Companion Proceedings of The 2019 World Wide Web Conference}} (San Francisco, USA) \emph{(\bibinfo{series}{WWW '19})}. \bibinfo{publisher}{Association for Computing Machinery}, \bibinfo{address}{New York, NY, USA}, \bibinfo{pages}{148–153}.
\newblock
\showISBNx{9781450366755}
\href{https://doi.org/10.1145/3308560.3316504}{doi:\nolinkurl{10.1145/3308560.3316504}}


\bibitem[Bankes(2002)]%
        {doi:10.1073/pnas.072081299}
\bibfield{author}{\bibinfo{person}{Steven~C. Bankes}.} \bibinfo{year}{2002}\natexlab{}.
\newblock \showarticletitle{Agent-based modeling: A revolution?}
\newblock \bibinfo{journal}{\emph{Proceedings of the National Academy of Sciences}} \bibinfo{volume}{99}, \bibinfo{number}{suppl\_3} (\bibinfo{year}{2002}), \bibinfo{pages}{7199--7200}.
\newblock
\href{https://doi.org/10.1073/pnas.072081299}{doi:\nolinkurl{10.1073/pnas.072081299}}
\showeprint{https://www.pnas.org/doi/pdf/10.1073/pnas.072081299}


\bibitem[Bates et~al\mbox{.}(1994)]%
        {bates1994role}
\bibfield{author}{\bibinfo{person}{Joseph Bates} {et~al\mbox{.}}} \bibinfo{year}{1994}\natexlab{}.
\newblock \showarticletitle{The role of emotion in believable agents}.
\newblock \bibinfo{journal}{\emph{Commun. ACM}} \bibinfo{volume}{37}, \bibinfo{number}{7} (\bibinfo{year}{1994}), \bibinfo{pages}{122--125}.
\newblock


\bibitem[Bommasani et~al\mbox{.}(2022)]%
        {bommasani2022opportunities}
\bibfield{author}{\bibinfo{person}{Rishi Bommasani}, \bibinfo{person}{Drew~A. Hudson}, \bibinfo{person}{Ehsan Adeli}, \bibinfo{person}{Russ Altman}, \bibinfo{person}{Simran Arora}, \bibinfo{person}{Sydney von Arx}, \bibinfo{person}{Michael~S. Bernstein}, \bibinfo{person}{Jeannette Bohg}, \bibinfo{person}{Antoine Bosselut}, \bibinfo{person}{Emma Brunskill}, \bibinfo{person}{Erik Brynjolfsson}, \bibinfo{person}{Shyamal Buch}, \bibinfo{person}{Dallas Card}, \bibinfo{person}{Rodrigo Castellon}, \bibinfo{person}{Niladri Chatterji}, \bibinfo{person}{Annie Chen}, \bibinfo{person}{Kathleen Creel}, \bibinfo{person}{Jared~Quincy Davis}, \bibinfo{person}{Dora Demszky}, \bibinfo{person}{Chris Donahue}, \bibinfo{person}{Moussa Doumbouya}, \bibinfo{person}{Esin Durmus}, \bibinfo{person}{Stefano Ermon}, \bibinfo{person}{John Etchemendy}, \bibinfo{person}{Kawin Ethayarajh}, \bibinfo{person}{Li Fei-Fei}, \bibinfo{person}{Chelsea Finn}, \bibinfo{person}{Trevor Gale}, \bibinfo{person}{Lauren Gillespie}, \bibinfo{person}{Karan
  Goel}, \bibinfo{person}{Noah Goodman}, \bibinfo{person}{Shelby Grossman}, \bibinfo{person}{Neel Guha}, \bibinfo{person}{Tatsunori Hashimoto}, \bibinfo{person}{Peter Henderson}, \bibinfo{person}{John Hewitt}, \bibinfo{person}{Daniel~E. Ho}, \bibinfo{person}{Jenny Hong}, \bibinfo{person}{Kyle Hsu}, \bibinfo{person}{Jing Huang}, \bibinfo{person}{Thomas Icard}, \bibinfo{person}{Saahil Jain}, \bibinfo{person}{Dan Jurafsky}, \bibinfo{person}{Pratyusha Kalluri}, \bibinfo{person}{Siddharth Karamcheti}, \bibinfo{person}{Geoff Keeling}, \bibinfo{person}{Fereshte Khani}, \bibinfo{person}{Omar Khattab}, \bibinfo{person}{Pang~Wei Koh}, \bibinfo{person}{Mark Krass}, \bibinfo{person}{Ranjay Krishna}, \bibinfo{person}{Rohith Kuditipudi}, \bibinfo{person}{Ananya Kumar}, \bibinfo{person}{Faisal Ladhak}, \bibinfo{person}{Mina Lee}, \bibinfo{person}{Tony Lee}, \bibinfo{person}{Jure Leskovec}, \bibinfo{person}{Isabelle Levent}, \bibinfo{person}{Xiang~Lisa Li}, \bibinfo{person}{Xuechen Li}, \bibinfo{person}{Tengyu Ma},
  \bibinfo{person}{Ali Malik}, \bibinfo{person}{Christopher~D. Manning}, \bibinfo{person}{Suvir Mirchandani}, \bibinfo{person}{Eric Mitchell}, \bibinfo{person}{Zanele Munyikwa}, \bibinfo{person}{Suraj Nair}, \bibinfo{person}{Avanika Narayan}, \bibinfo{person}{Deepak Narayanan}, \bibinfo{person}{Ben Newman}, \bibinfo{person}{Allen Nie}, \bibinfo{person}{Juan~Carlos Niebles}, \bibinfo{person}{Hamed Nilforoshan}, \bibinfo{person}{Julian Nyarko}, \bibinfo{person}{Giray Ogut}, \bibinfo{person}{Laurel Orr}, \bibinfo{person}{Isabel Papadimitriou}, \bibinfo{person}{Joon~Sung Park}, \bibinfo{person}{Chris Piech}, \bibinfo{person}{Eva Portelance}, \bibinfo{person}{Christopher Potts}, \bibinfo{person}{Aditi Raghunathan}, \bibinfo{person}{Rob Reich}, \bibinfo{person}{Hongyu Ren}, \bibinfo{person}{Frieda Rong}, \bibinfo{person}{Yusuf Roohani}, \bibinfo{person}{Camilo Ruiz}, \bibinfo{person}{Jack Ryan}, \bibinfo{person}{Christopher Ré}, \bibinfo{person}{Dorsa Sadigh}, \bibinfo{person}{Shiori Sagawa},
  \bibinfo{person}{Keshav Santhanam}, \bibinfo{person}{Andy Shih}, \bibinfo{person}{Krishnan Srinivasan}, \bibinfo{person}{Alex Tamkin}, \bibinfo{person}{Rohan Taori}, \bibinfo{person}{Armin~W. Thomas}, \bibinfo{person}{Florian Tramèr}, \bibinfo{person}{Rose~E. Wang}, \bibinfo{person}{William Wang}, \bibinfo{person}{Bohan Wu}, \bibinfo{person}{Jiajun Wu}, \bibinfo{person}{Yuhuai Wu}, \bibinfo{person}{Sang~Michael Xie}, \bibinfo{person}{Michihiro Yasunaga}, \bibinfo{person}{Jiaxuan You}, \bibinfo{person}{Matei Zaharia}, \bibinfo{person}{Michael Zhang}, \bibinfo{person}{Tianyi Zhang}, \bibinfo{person}{Xikun Zhang}, \bibinfo{person}{Yuhui Zhang}, \bibinfo{person}{Lucia Zheng}, \bibinfo{person}{Kaitlyn Zhou}, {and} \bibinfo{person}{Percy Liang}.} \bibinfo{year}{2022}\natexlab{}.
\newblock \bibinfo{title}{On the Opportunities and Risks of Foundation Models}.
\newblock
\showeprint[arxiv]{2108.07258}~[cs.LG]


\bibitem[Brenner(2010)]%
        {brenner_creating_2010}
\bibfield{author}{\bibinfo{person}{Michael Brenner}.} \bibinfo{year}{2010}\natexlab{}.
\newblock \showarticletitle{Creating {Dynamic} {Story} {Plots} with {Continual} {Multiagent} {Planning}}.
\newblock \bibinfo{journal}{\emph{Proceedings of the AAAI Conference on Artificial Intelligence}} \bibinfo{volume}{24}, \bibinfo{number}{1} (\bibinfo{date}{July} \bibinfo{year}{2010}), \bibinfo{pages}{1517--1522}.
\newblock
\href{https://doi.org/10.1609/aaai.v24i1.7567}{doi:\nolinkurl{10.1609/aaai.v24i1.7567}}


\bibitem[Brooks et~al\mbox{.}(1999)]%
        {10.1007/3-540-48834-0_5}
\bibfield{author}{\bibinfo{person}{Rodney~A. Brooks}, \bibinfo{person}{Cynthia Breazeal}, \bibinfo{person}{Matthew Marjanovi{\'{c}}}, \bibinfo{person}{Brian Scassellati}, {and} \bibinfo{person}{Matthew~M. Williamson}.} \bibinfo{year}{1999}\natexlab{}.
\newblock \showarticletitle{The Cog Project: Building a Humanoid Robot}. In \bibinfo{booktitle}{\emph{Computation for Metaphors, Analogy, and Agents}}, \bibfield{editor}{\bibinfo{person}{Chrystopher~L. Nehaniv}} (Ed.). \bibinfo{publisher}{Springer Berlin Heidelberg}, \bibinfo{address}{Berlin, Heidelberg}, \bibinfo{pages}{52--87}.
\newblock
\showISBNx{978-3-540-48834-7}


\bibitem[Brown et~al\mbox{.}(2020)]%
        {brown_language_2020}
\bibfield{author}{\bibinfo{person}{Tom Brown}, \bibinfo{person}{Benjamin Mann}, \bibinfo{person}{Nick Ryder}, \bibinfo{person}{Melanie Subbiah}, \bibinfo{person}{Jared~D Kaplan}, \bibinfo{person}{Prafulla Dhariwal}, \bibinfo{person}{Arvind Neelakantan}, \bibinfo{person}{Pranav Shyam}, \bibinfo{person}{Girish Sastry}, \bibinfo{person}{Amanda Askell}, \bibinfo{person}{Sandhini Agarwal}, \bibinfo{person}{Ariel Herbert-Voss}, \bibinfo{person}{Gretchen Krueger}, \bibinfo{person}{Tom Henighan}, \bibinfo{person}{Rewon Child}, \bibinfo{person}{Aditya Ramesh}, \bibinfo{person}{Daniel Ziegler}, \bibinfo{person}{Jeffrey Wu}, \bibinfo{person}{Clemens Winter}, \bibinfo{person}{Chris Hesse}, \bibinfo{person}{Mark Chen}, \bibinfo{person}{Eric Sigler}, \bibinfo{person}{Mateusz Litwin}, \bibinfo{person}{Scott Gray}, \bibinfo{person}{Benjamin Chess}, \bibinfo{person}{Jack Clark}, \bibinfo{person}{Christopher Berner}, \bibinfo{person}{Sam McCandlish}, \bibinfo{person}{Alec Radford}, \bibinfo{person}{Ilya Sutskever}, {and}
  \bibinfo{person}{Dario Amodei}.} \bibinfo{year}{2020}\natexlab{}.
\newblock \showarticletitle{Language {Models} are {Few}-{Shot} {Learners}}. In \bibinfo{booktitle}{\emph{Advances in {Neural} {Information} {Processing} {Systems}}}, \bibfield{editor}{\bibinfo{person}{H.~Larochelle}, \bibinfo{person}{M.~Ranzato}, \bibinfo{person}{R.~Hadsell}, \bibinfo{person}{M.~F. Balcan}, {and} \bibinfo{person}{H.~Lin}} (Eds.), Vol.~\bibinfo{volume}{33}. \bibinfo{publisher}{Curran Associates, Inc.}, \bibinfo{pages}{1877--1901}.
\newblock
\urldef\tempurl%
\url{https://proceedings.neurips.cc/paper_files/paper/2020/file/1457c0d6bfcb4967418bfb8ac142f64a-Paper.pdf}
\showURL{%
\tempurl}


\bibitem[Butler et~al\mbox{.}(2024)]%
        {butler_misinformation_2024}
\bibfield{author}{\bibinfo{person}{Lucy~H. Butler}, \bibinfo{person}{Padraig Lamont}, \bibinfo{person}{Dean Law~Yim Wan}, \bibinfo{person}{Toby Prike}, \bibinfo{person}{Mehwish Nasim}, \bibinfo{person}{Bradley Walker}, \bibinfo{person}{Nicolas Fay}, {and} \bibinfo{person}{Ullrich K.~H. Ecker}.} \bibinfo{year}{2024}\natexlab{}.
\newblock \showarticletitle{The ({Mis}){Information} {Game}: {A} social media simulator}.
\newblock \bibinfo{journal}{\emph{Behavior Research Methods}} \bibinfo{volume}{56}, \bibinfo{number}{3} (\bibinfo{date}{March} \bibinfo{year}{2024}), \bibinfo{pages}{2376--2397}.
\newblock
\showISSN{1554-3528}
\href{https://doi.org/10.3758/s13428-023-02153-x}{doi:\nolinkurl{10.3758/s13428-023-02153-x}}


\bibitem[Callison-Burch et~al\mbox{.}(2022)]%
        {callisonburch2022dungeons}
\bibfield{author}{\bibinfo{person}{Chris Callison-Burch}, \bibinfo{person}{Gaurav~Singh Tomar}, \bibinfo{person}{Lara~J. Martin}, \bibinfo{person}{Daphne Ippolito}, \bibinfo{person}{Suma Bailis}, {and} \bibinfo{person}{David Reitter}.} \bibinfo{year}{2022}\natexlab{}.
\newblock \bibinfo{title}{Dungeons and Dragons as a Dialog Challenge for Artificial Intelligence}.
\newblock
\showeprint[arxiv]{2210.07109}~[cs.CL]


\bibitem[Carley et~al\mbox{.}(2006)]%
        {1597399}
\bibfield{author}{\bibinfo{person}{K.M. Carley}, \bibinfo{person}{D.B. Fridsma}, \bibinfo{person}{E. Casman}, \bibinfo{person}{A. Yahja}, \bibinfo{person}{N. Altman}, \bibinfo{person}{Li-Chiou Chen}, \bibinfo{person}{B. Kaminsky}, {and} \bibinfo{person}{D. Nave}.} \bibinfo{year}{2006}\natexlab{}.
\newblock \showarticletitle{BioWar: scalable agent-based model of bioattacks}.
\newblock \bibinfo{journal}{\emph{IEEE Transactions on Systems, Man, and Cybernetics - Part A: Systems and Humans}} \bibinfo{volume}{36}, \bibinfo{number}{2} (\bibinfo{year}{2006}), \bibinfo{pages}{252--265}.
\newblock
\href{https://doi.org/10.1109/TSMCA.2005.851291}{doi:\nolinkurl{10.1109/TSMCA.2005.851291}}


\bibitem[Carr and Hayes(2015)]%
        {doi:10.1080/15456870.2015.972282}
\bibfield{author}{\bibinfo{person}{Caleb~T. Carr} {and} \bibinfo{person}{Rebecca~A. Hayes}.} \bibinfo{year}{2015}\natexlab{}.
\newblock \showarticletitle{Social Media: Defining, Developing, and Divining}.
\newblock \bibinfo{journal}{\emph{Atlantic Journal of Communication}} \bibinfo{volume}{23}, \bibinfo{number}{1} (\bibinfo{year}{2015}), \bibinfo{pages}{46--65}.
\newblock
\href{https://doi.org/10.1080/15456870.2015.972282}{doi:\nolinkurl{10.1080/15456870.2015.972282}}
\showeprint{https://doi.org/10.1080/15456870.2015.972282}


\bibitem[Conte and Paolucci(2014)]%
        {10.3389/fpsyg.2014.00668}
\bibfield{author}{\bibinfo{person}{Rosaria Conte} {and} \bibinfo{person}{Mario Paolucci}.} \bibinfo{year}{2014}\natexlab{}.
\newblock \showarticletitle{On agent-based modeling and computational social science}.
\newblock \bibinfo{journal}{\emph{Frontiers in Psychology}}  \bibinfo{volume}{5} (\bibinfo{year}{2014}).
\newblock
\showISSN{1664-1078}
\href{https://doi.org/10.3389/fpsyg.2014.00668}{doi:\nolinkurl{10.3389/fpsyg.2014.00668}}


\bibitem[Cresci(2020)]%
        {10.1145/3409116}
\bibfield{author}{\bibinfo{person}{Stefano Cresci}.} \bibinfo{year}{2020}\natexlab{}.
\newblock \showarticletitle{A Decade of Social Bot Detection}.
\newblock \bibinfo{journal}{\emph{Commun. ACM}} \bibinfo{volume}{63}, \bibinfo{number}{10} (\bibinfo{date}{sep} \bibinfo{year}{2020}), \bibinfo{pages}{72–83}.
\newblock
\showISSN{0001-0782}
\href{https://doi.org/10.1145/3409116}{doi:\nolinkurl{10.1145/3409116}}


\bibitem[danah boyd and Crawford(2012)]%
        {doi:10.1080/1369118X.2012.678878}
\bibfield{author}{\bibinfo{person}{danah boyd} {and} \bibinfo{person}{Kate Crawford}.} \bibinfo{year}{2012}\natexlab{}.
\newblock \showarticletitle{CRITICAL QUESTIONS FOR BIG DATA}.
\newblock \bibinfo{journal}{\emph{Information, Communication \& Society}} \bibinfo{volume}{15}, \bibinfo{number}{5} (\bibinfo{year}{2012}), \bibinfo{pages}{662--679}.
\newblock
\href{https://doi.org/10.1080/1369118X.2012.678878}{doi:\nolinkurl{10.1080/1369118X.2012.678878}}
\showeprint{https://doi.org/10.1080/1369118X.2012.678878}


\bibitem[Farmer and Axtell(2022)]%
        {axtell2022agent}
\bibfield{author}{\bibinfo{person}{J.~Doyne Farmer} {and} \bibinfo{person}{Robert~L. Axtell}.} \bibinfo{year}{2022}\natexlab{}.
\newblock \bibinfo{booktitle}{\emph{{Agent-Based Modeling in Economics and Finance: Past, Present, and Future}}}.
\newblock \bibinfo{type}{INET Oxford Working Papers} 2022-10. \bibinfo{institution}{Institute for New Economic Thinking at the Oxford Martin School, University of Oxford}.
\newblock
\urldef\tempurl%
\url{https://ideas.repec.org/p/amz/wpaper/2022-10.html}
\showURL{%
\tempurl}


\bibitem[Feng et~al\mbox{.}(2022a)]%
        {Feng_Tan_Li_Luo_2022}
\bibfield{author}{\bibinfo{person}{Shangbin Feng}, \bibinfo{person}{Zhaoxuan Tan}, \bibinfo{person}{Rui Li}, {and} \bibinfo{person}{Minnan Luo}.} \bibinfo{year}{2022}\natexlab{a}.
\newblock \showarticletitle{Heterogeneity-Aware Twitter Bot Detection with Relational Graph Transformers}.
\newblock \bibinfo{journal}{\emph{Proceedings of the AAAI Conference on Artificial Intelligence}} \bibinfo{volume}{36}, \bibinfo{number}{4} (\bibinfo{date}{Jun.} \bibinfo{year}{2022}), \bibinfo{pages}{3977--3985}.
\newblock
\href{https://doi.org/10.1609/aaai.v36i4.20314}{doi:\nolinkurl{10.1609/aaai.v36i4.20314}}


\bibitem[Feng et~al\mbox{.}(2022b)]%
        {feng_twibot-22_2022}
\bibfield{author}{\bibinfo{person}{Shangbin Feng}, \bibinfo{person}{Zhaoxuan Tan}, \bibinfo{person}{Herun Wan}, \bibinfo{person}{Ningnan Wang}, \bibinfo{person}{Zilong Chen}, \bibinfo{person}{Binchi Zhang}, \bibinfo{person}{Qinghua Zheng}, \bibinfo{person}{Wenqian Zhang}, \bibinfo{person}{Zhenyu Lei}, \bibinfo{person}{Shujie Yang}, \bibinfo{person}{Xinshun Feng}, \bibinfo{person}{Qingyue Zhang}, \bibinfo{person}{Hongrui Wang}, \bibinfo{person}{Yuhan Liu}, \bibinfo{person}{Yuyang Bai}, \bibinfo{person}{Heng Wang}, \bibinfo{person}{Zijian Cai}, \bibinfo{person}{Yanbo Wang}, \bibinfo{person}{Lijing Zheng}, \bibinfo{person}{Zihan Ma}, \bibinfo{person}{Jundong Li}, {and} \bibinfo{person}{Minnan Luo}.} \bibinfo{year}{2022}\natexlab{b}.
\newblock \showarticletitle{{TwiBot}-22: {Towards} {Graph}-{Based} {Twitter} {Bot} {Detection}}. In \bibinfo{booktitle}{\emph{Advances in {Neural} {Information} {Processing} {Systems}}}, \bibfield{editor}{\bibinfo{person}{S.~Koyejo}, \bibinfo{person}{S.~Mohamed}, \bibinfo{person}{A.~Agarwal}, \bibinfo{person}{D.~Belgrave}, \bibinfo{person}{K.~Cho}, {and} \bibinfo{person}{A.~Oh}} (Eds.), Vol.~\bibinfo{volume}{35}. \bibinfo{publisher}{Curran Associates, Inc.}, \bibinfo{pages}{35254--35269}.
\newblock
\urldef\tempurl%
\url{https://proceedings.neurips.cc/paper_files/paper/2022/file/e4fd610b1d77699a02df07ae97de992a-Paper-Datasets_and_Benchmarks.pdf}
\showURL{%
\tempurl}


\bibitem[Feng et~al\mbox{.}(2021)]%
        {10.1145/3459637.3481949}
\bibfield{author}{\bibinfo{person}{Shangbin Feng}, \bibinfo{person}{Herun Wan}, \bibinfo{person}{Ningnan Wang}, \bibinfo{person}{Jundong Li}, {and} \bibinfo{person}{Minnan Luo}.} \bibinfo{year}{2021}\natexlab{}.
\newblock \showarticletitle{SATAR: A Self-Supervised Approach to Twitter Account Representation Learning and Its Application in Bot Detection}. In \bibinfo{booktitle}{\emph{Proceedings of the 30th ACM International Conference on Information \& Knowledge Management}} (Virtual Event, Queensland, Australia) \emph{(\bibinfo{series}{CIKM '21})}. \bibinfo{publisher}{Association for Computing Machinery}, \bibinfo{address}{New York, NY, USA}, \bibinfo{pages}{3808–3817}.
\newblock
\showISBNx{9781450384469}
\href{https://doi.org/10.1145/3459637.3481949}{doi:\nolinkurl{10.1145/3459637.3481949}}


\bibitem[Franklin and Graesser(1997)]%
        {10.1007/BFb0013570}
\bibfield{author}{\bibinfo{person}{Stan Franklin} {and} \bibinfo{person}{Art Graesser}.} \bibinfo{year}{1997}\natexlab{}.
\newblock \showarticletitle{Is It an agent, or just a program?: A taxonomy for autonomous agents}. In \bibinfo{booktitle}{\emph{Intelligent Agents III Agent Theories, Architectures, and Languages}}, \bibfield{editor}{\bibinfo{person}{J{\"o}rg~P. M{\"u}ller}, \bibinfo{person}{Michael~J. Wooldridge}, {and} \bibinfo{person}{Nicholas~R. Jennings}} (Eds.). \bibinfo{publisher}{Springer Berlin Heidelberg}, \bibinfo{address}{Berlin, Heidelberg}, \bibinfo{pages}{21--35}.
\newblock
\showISBNx{978-3-540-68057-4}


\bibitem[Freiknecht and Effelsberg(2020)]%
        {10.1145/3402942.3409599}
\bibfield{author}{\bibinfo{person}{Jonas Freiknecht} {and} \bibinfo{person}{Wolfgang Effelsberg}.} \bibinfo{year}{2020}\natexlab{}.
\newblock \showarticletitle{Procedural Generation of Interactive Stories Using Language Models}. In \bibinfo{booktitle}{\emph{Proceedings of the 15th International Conference on the Foundations of Digital Games}} (Bugibba, Malta) \emph{(\bibinfo{series}{FDG '20})}. \bibinfo{publisher}{Association for Computing Machinery}, \bibinfo{address}{New York, NY, USA}, Article \bibinfo{articleno}{97}, \bibinfo{numpages}{8}~pages.
\newblock
\showISBNx{9781450388078}
\href{https://doi.org/10.1145/3402942.3409599}{doi:\nolinkurl{10.1145/3402942.3409599}}


\bibitem[Gao et~al\mbox{.}(2023a)]%
        {gao2023s3}
\bibfield{author}{\bibinfo{person}{Chen Gao}, \bibinfo{person}{Xiaochong Lan}, \bibinfo{person}{Zhihong Lu}, \bibinfo{person}{Jinzhu Mao}, \bibinfo{person}{Jinghua Piao}, \bibinfo{person}{Huandong Wang}, \bibinfo{person}{Depeng Jin}, {and} \bibinfo{person}{Yong Li}.} \bibinfo{year}{2023}\natexlab{a}.
\newblock \bibinfo{title}{S3: Social-network Simulation System with Large Language Model-Empowered Agents}.
\newblock
\showeprint[arxiv]{2307.14984}~[cs.SI]
\urldef\tempurl%
\url{https://arxiv.org/abs/2307.14984}
\showURL{%
\tempurl}


\bibitem[Gao et~al\mbox{.}(2023b)]%
        {10.1145/3568022}
\bibfield{author}{\bibinfo{person}{Chen Gao}, \bibinfo{person}{Yu Zheng}, \bibinfo{person}{Nian Li}, \bibinfo{person}{Yinfeng Li}, \bibinfo{person}{Yingrong Qin}, \bibinfo{person}{Jinghua Piao}, \bibinfo{person}{Yuhan Quan}, \bibinfo{person}{Jianxin Chang}, \bibinfo{person}{Depeng Jin}, \bibinfo{person}{Xiangnan He}, {and} \bibinfo{person}{Yong Li}.} \bibinfo{year}{2023}\natexlab{b}.
\newblock \showarticletitle{A Survey of Graph Neural Networks for Recommender Systems: Challenges, Methods, and Directions}.
\newblock \bibinfo{journal}{\emph{ACM Trans. Recomm. Syst.}} \bibinfo{volume}{1}, \bibinfo{number}{1}, Article \bibinfo{articleno}{3} (\bibinfo{date}{mar} \bibinfo{year}{2023}), \bibinfo{numpages}{51}~pages.
\newblock
\href{https://doi.org/10.1145/3568022}{doi:\nolinkurl{10.1145/3568022}}


\bibitem[Gatti et~al\mbox{.}(2014)]%
        {10.1007/978-3-642-54783-6_2}
\bibfield{author}{\bibinfo{person}{Ma{\'i}ra Gatti}, \bibinfo{person}{Paulo Cavalin}, \bibinfo{person}{Samuel~Barbosa Neto}, \bibinfo{person}{Claudio Pinhanez}, \bibinfo{person}{C{\'i}cero dos Santos}, \bibinfo{person}{Daniel Gribel}, {and} \bibinfo{person}{Ana~Paula Appel}.} \bibinfo{year}{2014}\natexlab{}.
\newblock \showarticletitle{Large-Scale Multi-agent-Based Modeling and Simulation of Microblogging-Based Online Social Network}. In \bibinfo{booktitle}{\emph{Multi-Agent-Based Simulation XIV}}, \bibfield{editor}{\bibinfo{person}{Shah~Jamal Alam} {and} \bibinfo{person}{H.~Van~Dyke Parunak}} (Eds.). \bibinfo{publisher}{Springer Berlin Heidelberg}, \bibinfo{address}{Berlin, Heidelberg}, \bibinfo{pages}{17--33}.
\newblock
\showISBNx{978-3-642-54783-6}


\bibitem[Ge et~al\mbox{.}(2013)]%
        {doi:10.1177/0037549713477682}
\bibfield{author}{\bibinfo{person}{Yuanzheng Ge}, \bibinfo{person}{Liang Liu}, \bibinfo{person}{Xiaogang Qiu}, \bibinfo{person}{Hongbin Song}, \bibinfo{person}{Yong Wang}, {and} \bibinfo{person}{Kedi Huang}.} \bibinfo{year}{2013}\natexlab{}.
\newblock \showarticletitle{A framework of multilayer social networks for communication behavior with agent-based modeling}.
\newblock \bibinfo{journal}{\emph{SIMULATION}} \bibinfo{volume}{89}, \bibinfo{number}{7} (\bibinfo{year}{2013}), \bibinfo{pages}{810--828}.
\newblock
\href{https://doi.org/10.1177/0037549713477682}{doi:\nolinkurl{10.1177/0037549713477682}}
\showeprint{https://doi.org/10.1177/0037549713477682}


\bibitem[Ghani et~al\mbox{.}(2019)]%
        {GHANI2019417}
\bibfield{author}{\bibinfo{person}{Norjihan~Abdul Ghani}, \bibinfo{person}{Suraya Hamid}, \bibinfo{person}{Ibrahim~Abaker {Targio Hashem}}, {and} \bibinfo{person}{Ejaz Ahmed}.} \bibinfo{year}{2019}\natexlab{}.
\newblock \showarticletitle{Social media big data analytics: A survey}.
\newblock \bibinfo{journal}{\emph{Computers in Human Behavior}}  \bibinfo{volume}{101} (\bibinfo{year}{2019}), \bibinfo{pages}{417--428}.
\newblock
\showISSN{0747-5632}
\href{https://doi.org/10.1016/j.chb.2018.08.039}{doi:\nolinkurl{10.1016/j.chb.2018.08.039}}


\bibitem[H\"{a}m\"{a}l\"{a}inen et~al\mbox{.}(2023)]%
        {10.1145/3544548.3580688}
\bibfield{author}{\bibinfo{person}{Perttu H\"{a}m\"{a}l\"{a}inen}, \bibinfo{person}{Mikke Tavast}, {and} \bibinfo{person}{Anton Kunnari}.} \bibinfo{year}{2023}\natexlab{}.
\newblock \showarticletitle{Evaluating Large Language Models in Generating Synthetic HCI Research Data: A Case Study}. In \bibinfo{booktitle}{\emph{Proceedings of the 2023 CHI Conference on Human Factors in Computing Systems}} (Hamburg, Germany) \emph{(\bibinfo{series}{CHI '23})}. \bibinfo{publisher}{Association for Computing Machinery}, \bibinfo{address}{New York, NY, USA}, Article \bibinfo{articleno}{433}, \bibinfo{numpages}{19}~pages.
\newblock
\showISBNx{9781450394215}
\href{https://doi.org/10.1145/3544548.3580688}{doi:\nolinkurl{10.1145/3544548.3580688}}


\bibitem[Helbing(2012)]%
        {helbing_agent-based_2012}
\bibfield{author}{\bibinfo{person}{Dirk Helbing}.} \bibinfo{year}{2012}\natexlab{}.
\newblock \showarticletitle{Agent-{Based} {Modeling}}.
\newblock In \bibinfo{booktitle}{\emph{Social {Self}-{Organization}: {Agent}-{Based} {Simulations} and {Experiments} to {Study} {Emergent} {Social} {Behavior}}}, \bibfield{editor}{\bibinfo{person}{Dirk Helbing}} (Ed.). \bibinfo{publisher}{Springer}, \bibinfo{address}{Berlin, Heidelberg}, \bibinfo{pages}{25--70}.
\newblock
\showISBNx{978-3-642-24004-1}
\href{https://doi.org/10.1007/978-3-642-24004-1_2}{doi:\nolinkurl{10.1007/978-3-642-24004-1_2}}


\bibitem[Holyoak and Morrison(2005)]%
        {holyoak2005cambridge}
\bibfield{author}{\bibinfo{person}{Keith~J Holyoak} {and} \bibinfo{person}{Robert~G Morrison}.} \bibinfo{year}{2005}\natexlab{}.
\newblock \bibinfo{booktitle}{\emph{The Cambridge handbook of thinking and reasoning}}.
\newblock \bibinfo{publisher}{Cambridge University Press}.
\newblock


\bibitem[Hong et~al\mbox{.}(2023)]%
        {hong2023metagptmetaprogrammingmultiagent}
\bibfield{author}{\bibinfo{person}{Sirui Hong}, \bibinfo{person}{Mingchen Zhuge}, \bibinfo{person}{Jonathan Chen}, \bibinfo{person}{Xiawu Zheng}, \bibinfo{person}{Yuheng Cheng}, \bibinfo{person}{Ceyao Zhang}, \bibinfo{person}{Jinlin Wang}, \bibinfo{person}{Zili Wang}, \bibinfo{person}{Steven Ka~Shing Yau}, \bibinfo{person}{Zijuan Lin}, \bibinfo{person}{Liyang Zhou}, \bibinfo{person}{Chenyu Ran}, \bibinfo{person}{Lingfeng Xiao}, \bibinfo{person}{Chenglin Wu}, {and} \bibinfo{person}{Jürgen Schmidhuber}.} \bibinfo{year}{2023}\natexlab{}.
\newblock \bibinfo{title}{MetaGPT: Meta Programming for A Multi-Agent Collaborative Framework}.
\newblock
\showeprint[arxiv]{2308.00352}~[cs.AI]
\urldef\tempurl%
\url{https://arxiv.org/abs/2308.00352}
\showURL{%
\tempurl}


\bibitem[Horton(2023)]%
        {NBERw31122}
\bibfield{author}{\bibinfo{person}{John~J Horton}.} \bibinfo{year}{2023}\natexlab{}.
\newblock \bibinfo{booktitle}{\emph{Large Language Models as Simulated Economic Agents: What Can We Learn from Homo Silicus?}}
\newblock \bibinfo{type}{Working Paper} 31122. \bibinfo{institution}{National Bureau of Economic Research}.
\newblock
\href{https://doi.org/10.3386/w31122}{doi:\nolinkurl{10.3386/w31122}}


\bibitem[ISBISTER and NASS(2000)]%
        {isbister_consistency_2000}
\bibfield{author}{\bibinfo{person}{KATHERINE ISBISTER} {and} \bibinfo{person}{CLIFFORD NASS}.} \bibinfo{year}{2000}\natexlab{}.
\newblock \showarticletitle{Consistency of personality in interactive characters: verbal cues, non-verbal cues, and user characteristics}.
\newblock \bibinfo{journal}{\emph{International Journal of Human-Computer Studies}} \bibinfo{volume}{53}, \bibinfo{number}{2} (\bibinfo{year}{2000}), \bibinfo{pages}{251--267}.
\newblock
\showISSN{1071-5819}
\href{https://doi.org/10.1006/ijhc.2000.0368}{doi:\nolinkurl{10.1006/ijhc.2000.0368}}


\bibitem[Kaiya et~al\mbox{.}(2023)]%
        {kaiya2023lyfeagentsgenerativeagents}
\bibfield{author}{\bibinfo{person}{Zhao Kaiya}, \bibinfo{person}{Michelangelo Naim}, \bibinfo{person}{Jovana Kondic}, \bibinfo{person}{Manuel Cortes}, \bibinfo{person}{Jiaxin Ge}, \bibinfo{person}{Shuying Luo}, \bibinfo{person}{Guangyu~Robert Yang}, {and} \bibinfo{person}{Andrew Ahn}.} \bibinfo{year}{2023}\natexlab{}.
\newblock \bibinfo{title}{Lyfe Agents: Generative agents for low-cost real-time social interactions}.
\newblock
\showeprint[arxiv]{2310.02172}~[cs.HC]
\urldef\tempurl%
\url{https://arxiv.org/abs/2310.02172}
\showURL{%
\tempurl}


\bibitem[Kaligotla et~al\mbox{.}(2015)]%
        {7408553}
\bibfield{author}{\bibinfo{person}{Chaitanya Kaligotla}, \bibinfo{person}{Enver Yücesan}, {and} \bibinfo{person}{Stephen~E. Chick}.} \bibinfo{year}{2015}\natexlab{}.
\newblock \showarticletitle{An agent based model of spread of competing rumors through online interactions on social media}. In \bibinfo{booktitle}{\emph{2015 Winter Simulation Conference (WSC)}}. \bibinfo{pages}{3985--3996}.
\newblock
\href{https://doi.org/10.1109/WSC.2015.7408553}{doi:\nolinkurl{10.1109/WSC.2015.7408553}}


\bibitem[Kaplan and Haenlein(2010)]%
        {KAPLAN201059}
\bibfield{author}{\bibinfo{person}{Andreas~M. Kaplan} {and} \bibinfo{person}{Michael Haenlein}.} \bibinfo{year}{2010}\natexlab{}.
\newblock \showarticletitle{Users of the world, unite! The challenges and opportunities of Social Media}.
\newblock \bibinfo{journal}{\emph{Business Horizons}} \bibinfo{volume}{53}, \bibinfo{number}{1} (\bibinfo{year}{2010}), \bibinfo{pages}{59--68}.
\newblock
\showISSN{0007-6813}
\href{https://doi.org/10.1016/j.bushor.2009.09.003}{doi:\nolinkurl{10.1016/j.bushor.2009.09.003}}


\bibitem[Kapoor et~al\mbox{.}(2018)]%
        {kapoor_advances_2018}
\bibfield{author}{\bibinfo{person}{Kawaljeet~Kaur Kapoor}, \bibinfo{person}{Kuttimani Tamilmani}, \bibinfo{person}{Nripendra~P. Rana}, \bibinfo{person}{Pushp Patil}, \bibinfo{person}{Yogesh~K. Dwivedi}, {and} \bibinfo{person}{Sridhar Nerur}.} \bibinfo{year}{2018}\natexlab{}.
\newblock \showarticletitle{Advances in {Social} {Media} {Research}: {Past}, {Present} and {Future}}.
\newblock \bibinfo{journal}{\emph{Information Systems Frontiers}} \bibinfo{volume}{20}, \bibinfo{number}{3} (\bibinfo{date}{June} \bibinfo{year}{2018}), \bibinfo{pages}{531--558}.
\newblock
\showISSN{1572-9419}
\href{https://doi.org/10.1007/s10796-017-9810-y}{doi:\nolinkurl{10.1007/s10796-017-9810-y}}


\bibitem[Kerr et~al\mbox{.}(2021)]%
        {kerr_covasim_2021}
\bibfield{author}{\bibinfo{person}{Cliff~C. Kerr}, \bibinfo{person}{Robyn~M. Stuart}, \bibinfo{person}{Dina Mistry}, \bibinfo{person}{Romesh~G. Abeysuriya}, \bibinfo{person}{Katherine Rosenfeld}, \bibinfo{person}{Gregory~R. Hart}, \bibinfo{person}{Rafael~C. Núñez}, \bibinfo{person}{Jamie~A. Cohen}, \bibinfo{person}{Prashanth Selvaraj}, \bibinfo{person}{Brittany Hagedorn}, \bibinfo{person}{Lauren George}, \bibinfo{person}{Michał Jastrzębski}, \bibinfo{person}{Amanda~S. Izzo}, \bibinfo{person}{Greer Fowler}, \bibinfo{person}{Anna Palmer}, \bibinfo{person}{Dominic Delport}, \bibinfo{person}{Nick Scott}, \bibinfo{person}{Sherrie~L. Kelly}, \bibinfo{person}{Caroline~S. Bennette}, \bibinfo{person}{Bradley~G. Wagner}, \bibinfo{person}{Stewart~T. Chang}, \bibinfo{person}{Assaf~P. Oron}, \bibinfo{person}{Edward~A. Wenger}, \bibinfo{person}{Jasmina Panovska-Griffiths}, \bibinfo{person}{Michael Famulare}, {and} \bibinfo{person}{Daniel~J. Klein}.} \bibinfo{year}{2021}\natexlab{}.
\newblock \showarticletitle{Covasim: {An} agent-based model of {COVID}-19 dynamics and interventions}.
\newblock \bibinfo{journal}{\emph{PLOS Computational Biology}} \bibinfo{volume}{17}, \bibinfo{number}{7} (\bibinfo{date}{July} \bibinfo{year}{2021}), \bibinfo{pages}{e1009149}.
\newblock
\showISSN{1553-7358}
\href{https://doi.org/10.1371/journal.pcbi.1009149}{doi:\nolinkurl{10.1371/journal.pcbi.1009149}}
\newblock
\shownote{Publisher: Public Library of Science}.


\bibitem[Kudugunta and Ferrara(2018)]%
        {kudugunta_deep_2018}
\bibfield{author}{\bibinfo{person}{Sneha Kudugunta} {and} \bibinfo{person}{Emilio Ferrara}.} \bibinfo{year}{2018}\natexlab{}.
\newblock \showarticletitle{Deep neural networks for bot detection}.
\newblock \bibinfo{journal}{\emph{Information Sciences}}  \bibinfo{volume}{467} (\bibinfo{year}{2018}), \bibinfo{pages}{312--322}.
\newblock
\showISSN{0020-0255}
\href{https://doi.org/10.1016/j.ins.2018.08.019}{doi:\nolinkurl{10.1016/j.ins.2018.08.019}}


\bibitem[Laird and VanLent(2001)]%
        {laird_human-level_2001}
\bibfield{author}{\bibinfo{person}{John Laird} {and} \bibinfo{person}{Michael VanLent}.} \bibinfo{year}{2001}\natexlab{}.
\newblock \showarticletitle{Human-{Level} {AI}’s {Killer} {Application}: {Interactive} {Computer} {Games}}.
\newblock \bibinfo{journal}{\emph{AI Magazine}} \bibinfo{volume}{22}, \bibinfo{number}{2} (\bibinfo{date}{June} \bibinfo{year}{2001}), \bibinfo{pages}{15}.
\newblock
\href{https://doi.org/10.1609/aimag.v22i2.1558}{doi:\nolinkurl{10.1609/aimag.v22i2.1558}}


\bibitem[Laird(2001)]%
        {10.1145/375735.376343}
\bibfield{author}{\bibinfo{person}{John~E. Laird}.} \bibinfo{year}{2001}\natexlab{}.
\newblock \showarticletitle{It Knows What You're Going to Do: Adding Anticipation to a Quakebot}. In \bibinfo{booktitle}{\emph{Proceedings of the Fifth International Conference on Autonomous Agents}} (Montreal, Quebec, Canada) \emph{(\bibinfo{series}{AGENTS '01})}. \bibinfo{publisher}{Association for Computing Machinery}, \bibinfo{address}{New York, NY, USA}, \bibinfo{pages}{385–392}.
\newblock
\showISBNx{158113326X}
\href{https://doi.org/10.1145/375735.376343}{doi:\nolinkurl{10.1145/375735.376343}}


\bibitem[Leong et~al\mbox{.}(2015)]%
        {leong_ict-enabled_2015}
\bibfield{author}{\bibinfo{person}{Carmen Leong}, \bibinfo{person}{Shan Pan}, \bibinfo{person}{Peter Ractham}, {and} \bibinfo{person}{Laddawan Kaewkitipong}.} \bibinfo{year}{2015}\natexlab{}.
\newblock \showarticletitle{{ICT}-{Enabled} {Community} {Empowerment} in {Crisis} {Response}: {Social} {Media} in {Thailand} {Flooding} 2011}.
\newblock \bibinfo{journal}{\emph{Journal of the Association for Information Systems}} \bibinfo{volume}{16}, \bibinfo{number}{3} (\bibinfo{date}{March} \bibinfo{year}{2015}).
\newblock
\showISSN{1536-9323}
\href{https://doi.org/10.17705/1jais.00390}{doi:\nolinkurl{10.17705/1jais.00390}}


\bibitem[Li et~al\mbox{.}(2023)]%
        {li2023masqueradeexploringbehaviorimpact}
\bibfield{author}{\bibinfo{person}{Siyu Li}, \bibinfo{person}{Jin Yang}, {and} \bibinfo{person}{Kui Zhao}.} \bibinfo{year}{2023}\natexlab{}.
\newblock \bibinfo{title}{Are you in a Masquerade? Exploring the Behavior and Impact of Large Language Model Driven Social Bots in Online Social Networks}.
\newblock
\showeprint[arxiv]{2307.10337}~[cs.SI]
\urldef\tempurl%
\url{https://arxiv.org/abs/2307.10337}
\showURL{%
\tempurl}


\bibitem[Lin et~al\mbox{.}(2023)]%
        {lin2023agentsimsopensourcesandboxlarge}
\bibfield{author}{\bibinfo{person}{Jiaju Lin}, \bibinfo{person}{Haoran Zhao}, \bibinfo{person}{Aochi Zhang}, \bibinfo{person}{Yiting Wu}, \bibinfo{person}{Huqiuyue Ping}, {and} \bibinfo{person}{Qin Chen}.} \bibinfo{year}{2023}\natexlab{}.
\newblock \bibinfo{title}{AgentSims: An Open-Source Sandbox for Large Language Model Evaluation}.
\newblock
\showeprint[arxiv]{2308.04026}~[cs.AI]
\urldef\tempurl%
\url{https://arxiv.org/abs/2308.04026}
\showURL{%
\tempurl}


\bibitem[Lundmark et~al\mbox{.}(2017)]%
        {lundmark_little_2017}
\bibfield{author}{\bibinfo{person}{Leif~W. Lundmark}, \bibinfo{person}{Chong Oh}, {and} \bibinfo{person}{J.~Cameron Verhaal}.} \bibinfo{year}{2017}\natexlab{}.
\newblock \showarticletitle{A little {Birdie} told me: {Social} media, organizational legitimacy, and underpricing in initial public offerings}.
\newblock \bibinfo{journal}{\emph{Information Systems Frontiers}} \bibinfo{volume}{19}, \bibinfo{number}{6} (\bibinfo{date}{Dec.} \bibinfo{year}{2017}), \bibinfo{pages}{1407--1422}.
\newblock
\showISSN{1572-9419}
\href{https://doi.org/10.1007/s10796-016-9654-x}{doi:\nolinkurl{10.1007/s10796-016-9654-x}}


\bibitem[Macal and North(2005)]%
        {1574234}
\bibfield{author}{\bibinfo{person}{C.M. Macal} {and} \bibinfo{person}{M.J. North}.} \bibinfo{year}{2005}\natexlab{}.
\newblock \showarticletitle{Tutorial on agent-based modeling and simulation}. In \bibinfo{booktitle}{\emph{Proceedings of the Winter Simulation Conference, 2005.}} \bibinfo{pages}{14 pp.--}.
\newblock
\href{https://doi.org/10.1109/WSC.2005.1574234}{doi:\nolinkurl{10.1109/WSC.2005.1574234}}


\bibitem[Macal and North(2009)]%
        {5429318}
\bibfield{author}{\bibinfo{person}{Charles~M. Macal} {and} \bibinfo{person}{Michael~J. North}.} \bibinfo{year}{2009}\natexlab{}.
\newblock \showarticletitle{Agent-based modeling and simulation}. In \bibinfo{booktitle}{\emph{Proceedings of the 2009 Winter Simulation Conference (WSC)}}. \bibinfo{pages}{86--98}.
\newblock
\href{https://doi.org/10.1109/WSC.2009.5429318}{doi:\nolinkurl{10.1109/WSC.2009.5429318}}


\bibitem[Marcotte and Hamilton(2017)]%
        {marcotte_behavior_2017}
\bibfield{author}{\bibinfo{person}{Ryan Marcotte} {and} \bibinfo{person}{Howard~J. Hamilton}.} \bibinfo{year}{2017}\natexlab{}.
\newblock \showarticletitle{Behavior {Trees} for {Modelling} {Artificial} {Intelligence} in {Games}: {A} {Tutorial}}.
\newblock \bibinfo{journal}{\emph{The Computer Games Journal}} \bibinfo{volume}{6}, \bibinfo{number}{3} (\bibinfo{date}{Sept.} \bibinfo{year}{2017}), \bibinfo{pages}{171--184}.
\newblock
\showISSN{2052-773X}
\href{https://doi.org/10.1007/s40869-017-0040-9}{doi:\nolinkurl{10.1007/s40869-017-0040-9}}


\bibitem[Markel et~al\mbox{.}(2023)]%
        {markel2023gpteach}
\bibfield{author}{\bibinfo{person}{Julia~M. Markel}, \bibinfo{person}{Steven~G. Opferman}, \bibinfo{person}{James~A. Landay}, {and} \bibinfo{person}{Chris Piech}.} \bibinfo{year}{2023}\natexlab{}.
\newblock \showarticletitle{GPTeach: Interactive TA Training with GPT-based Students}. In \bibinfo{booktitle}{\emph{Proceedings of the Tenth ACM Conference on Learning @ Scale}} (Copenhagen, Denmark) \emph{(\bibinfo{series}{L@S '23})}. \bibinfo{publisher}{Association for Computing Machinery}, \bibinfo{address}{New York, NY, USA}, \bibinfo{pages}{226–236}.
\newblock
\showISBNx{9798400700255}
\href{https://doi.org/10.1145/3573051.3593393}{doi:\nolinkurl{10.1145/3573051.3593393}}


\bibitem[Miyashita et~al\mbox{.}(2017)]%
        {8022767}
\bibfield{author}{\bibinfo{person}{Shohei Miyashita}, \bibinfo{person}{Xinyu Lian}, \bibinfo{person}{Xiao Zeng}, \bibinfo{person}{Takashi Matsubara}, {and} \bibinfo{person}{Kuniaki Uehara}.} \bibinfo{year}{2017}\natexlab{}.
\newblock \showarticletitle{Developing game AI agent behaving like human by mixing reinforcement learning and supervised learning}. In \bibinfo{booktitle}{\emph{2017 18th IEEE/ACIS International Conference on Software Engineering, Artificial Intelligence, Networking and Parallel/Distributed Computing (SNPD)}}. \bibinfo{pages}{489--494}.
\newblock
\href{https://doi.org/10.1109/SNPD.2017.8022767}{doi:\nolinkurl{10.1109/SNPD.2017.8022767}}


\bibitem[Moreno et~al\mbox{.}(2013)]%
        {doi:10.1089/cyber.2012.0334}
\bibfield{author}{\bibinfo{person}{Megan~A. Moreno}, \bibinfo{person}{Natalie Goniu}, \bibinfo{person}{Peter~S. Moreno}, {and} \bibinfo{person}{Douglas Diekema}.} \bibinfo{year}{2013}\natexlab{}.
\newblock \showarticletitle{Ethics of Social Media Research: Common Concerns and Practical Considerations}.
\newblock \bibinfo{journal}{\emph{Cyberpsychology, Behavior, and Social Networking}} \bibinfo{volume}{16}, \bibinfo{number}{9} (\bibinfo{year}{2013}), \bibinfo{pages}{708--713}.
\newblock
\href{https://doi.org/10.1089/cyber.2012.0334}{doi:\nolinkurl{10.1089/cyber.2012.0334}}
\showeprint{https://doi.org/10.1089/cyber.2012.0334}
\newblock
\shownote{PMID: 23679571}.


\bibitem[Myers(2019)]%
        {noauthor_qualitative_2019}
\bibfield{author}{\bibinfo{person}{Michael~D Myers}.} \bibinfo{year}{2019}\natexlab{}.
\newblock \bibinfo{booktitle}{\emph{Qualitative {Research} in {Business} and {Management}}}.
\newblock \bibinfo{publisher}{SAGE Publications Ltd}, \bibinfo{address}{London}.
\newblock
\showISBNx{978-1-5264-1832-6}
\urldef\tempurl%
\url{http://digital.casalini.it/9781526418326}
\showURL{%
\tempurl}


\bibitem[Nakano et~al\mbox{.}(2022)]%
        {nakano2022webgptbrowserassistedquestionansweringhuman}
\bibfield{author}{\bibinfo{person}{Reiichiro Nakano}, \bibinfo{person}{Jacob Hilton}, \bibinfo{person}{Suchir Balaji}, \bibinfo{person}{Jeff Wu}, \bibinfo{person}{Long Ouyang}, \bibinfo{person}{Christina Kim}, \bibinfo{person}{Christopher Hesse}, \bibinfo{person}{Shantanu Jain}, \bibinfo{person}{Vineet Kosaraju}, \bibinfo{person}{William Saunders}, \bibinfo{person}{Xu Jiang}, \bibinfo{person}{Karl Cobbe}, \bibinfo{person}{Tyna Eloundou}, \bibinfo{person}{Gretchen Krueger}, \bibinfo{person}{Kevin Button}, \bibinfo{person}{Matthew Knight}, \bibinfo{person}{Benjamin Chess}, {and} \bibinfo{person}{John Schulman}.} \bibinfo{year}{2022}\natexlab{}.
\newblock \bibinfo{title}{WebGPT: Browser-assisted question-answering with human feedback}.
\newblock
\showeprint[arxiv]{2112.09332}~[cs.CL]
\urldef\tempurl%
\url{https://arxiv.org/abs/2112.09332}
\showURL{%
\tempurl}


\bibitem[Onuchowska and Berndt(2019)]%
        {onuchowska2019using}
\bibfield{author}{\bibinfo{person}{Agnieszka Onuchowska} {and} \bibinfo{person}{Donald~J Berndt}.} \bibinfo{year}{2019}\natexlab{}.
\newblock \showarticletitle{Using Agent-Based Modelling to Address Malicious Behavior on Social Media.}. In \bibinfo{booktitle}{\emph{ICIS}}.
\newblock
\urldef\tempurl%
\url{https://core.ac.uk/download/pdf/301383842.pdf}
\showURL{%
\tempurl}


\bibitem[Park et~al\mbox{.}(2023)]%
        {park2023generative}
\bibfield{author}{\bibinfo{person}{Joon~Sung Park}, \bibinfo{person}{Joseph~C. O'Brien}, \bibinfo{person}{Carrie~J. Cai}, \bibinfo{person}{Meredith~Ringel Morris}, \bibinfo{person}{Percy Liang}, {and} \bibinfo{person}{Michael~S. Bernstein}.} \bibinfo{year}{2023}\natexlab{}.
\newblock \bibinfo{title}{Generative Agents: Interactive Simulacra of Human Behavior}.
\newblock
\showeprint[arxiv]{2304.03442}~[cs.HC]


\bibitem[Park et~al\mbox{.}(2022)]%
        {10.1145/3526113.3545616}
\bibfield{author}{\bibinfo{person}{Joon~Sung Park}, \bibinfo{person}{Lindsay Popowski}, \bibinfo{person}{Carrie Cai}, \bibinfo{person}{Meredith~Ringel Morris}, \bibinfo{person}{Percy Liang}, {and} \bibinfo{person}{Michael~S. Bernstein}.} \bibinfo{year}{2022}\natexlab{}.
\newblock \showarticletitle{Social Simulacra: Creating Populated Prototypes for Social Computing Systems}. In \bibinfo{booktitle}{\emph{Proceedings of the 35th Annual ACM Symposium on User Interface Software and Technology}} (Bend, OR, USA) \emph{(\bibinfo{series}{UIST '22})}. \bibinfo{publisher}{Association for Computing Machinery}, \bibinfo{address}{New York, NY, USA}, Article \bibinfo{articleno}{74}, \bibinfo{numpages}{18}~pages.
\newblock
\showISBNx{9781450393201}
\href{https://doi.org/10.1145/3526113.3545616}{doi:\nolinkurl{10.1145/3526113.3545616}}


\bibitem[Riedl and Bulitko(2021)]%
        {riedl_interactive_2021}
\bibfield{author}{\bibinfo{person}{Mark Riedl} {and} \bibinfo{person}{Vadim Bulitko}.} \bibinfo{year}{2021}\natexlab{}.
\newblock \showarticletitle{Interactive {Narrative}: {A} {Novel} {Application} of {Artificial} {Intelligence} for {Computer} {Games}}.
\newblock \bibinfo{journal}{\emph{Proceedings of the AAAI Conference on Artificial Intelligence}} \bibinfo{volume}{26}, \bibinfo{number}{1} (\bibinfo{date}{Sept.} \bibinfo{year}{2021}), \bibinfo{pages}{2160--2165}.
\newblock
\href{https://doi.org/10.1609/aaai.v26i1.8447}{doi:\nolinkurl{10.1609/aaai.v26i1.8447}}


\bibitem[Rips and Conrad(1989)]%
        {rips_folk_1989}
\bibfield{author}{\bibinfo{person}{Lance~J. Rips} {and} \bibinfo{person}{Frederick~G. Conrad}.} \bibinfo{year}{1989}\natexlab{}.
\newblock \showarticletitle{Folk psychology of mental activities}.
\newblock \bibinfo{journal}{\emph{Psychological Review}} \bibinfo{volume}{96}, \bibinfo{number}{2} (\bibinfo{year}{1989}), \bibinfo{pages}{187--207}.
\newblock
\showISSN{1939-1471}
\href{https://doi.org/10.1037/0033-295X.96.2.187}{doi:\nolinkurl{10.1037/0033-295X.96.2.187}}
\newblock
\shownote{Place: US Publisher: American Psychological Association}.


\bibitem[Russell and Norvig(2016)]%
        {russell2016artificial}
\bibfield{author}{\bibinfo{person}{Stuart~J Russell} {and} \bibinfo{person}{Peter Norvig}.} \bibinfo{year}{2016}\natexlab{}.
\newblock \bibinfo{booktitle}{\emph{Artificial intelligence: a modern approach}}.
\newblock \bibinfo{publisher}{Pearson}.
\newblock


\bibitem[Sathanur et~al\mbox{.}(2015)]%
        {7225278}
\bibfield{author}{\bibinfo{person}{Arun~V Sathanur}, \bibinfo{person}{Miao Sui}, {and} \bibinfo{person}{Vikram Jandhyala}.} \bibinfo{year}{2015}\natexlab{}.
\newblock \showarticletitle{Assessing strategies for controlling viral rumor propagation on social media - a simulation approach}. In \bibinfo{booktitle}{\emph{2015 IEEE International Symposium on Technologies for Homeland Security (HST)}}. \bibinfo{pages}{1--6}.
\newblock
\href{https://doi.org/10.1109/THS.2015.7225278}{doi:\nolinkurl{10.1109/THS.2015.7225278}}


\bibitem[Savaglio et~al\mbox{.}(2020)]%
        {savaglio_agent-based_2020}
\bibfield{author}{\bibinfo{person}{Claudio Savaglio}, \bibinfo{person}{Maria Ganzha}, \bibinfo{person}{Marcin Paprzycki}, \bibinfo{person}{Costin Bădică}, \bibinfo{person}{Mirjana Ivanović}, {and} \bibinfo{person}{Giancarlo Fortino}.} \bibinfo{year}{2020}\natexlab{}.
\newblock \showarticletitle{Agent-based {Internet} of {Things}: {State}-of-the-art and research challenges}.
\newblock \bibinfo{journal}{\emph{Future Generation Computer Systems}}  \bibinfo{volume}{102} (\bibinfo{date}{Jan.} \bibinfo{year}{2020}), \bibinfo{pages}{1038--1053}.
\newblock
\showISSN{0167739X}
\href{https://doi.org/10.1016/j.future.2019.09.016}{doi:\nolinkurl{10.1016/j.future.2019.09.016}}


\bibitem[Scher and Schett(2021)]%
        {scher2021key}
\bibfield{author}{\bibinfo{person}{Jose~U Scher} {and} \bibinfo{person}{Georg Schett}.} \bibinfo{year}{2021}\natexlab{}.
\newblock \showarticletitle{Key opinion leaders—a critical perspective}.
\newblock \bibinfo{journal}{\emph{Nature Reviews Rheumatology}} \bibinfo{volume}{17}, \bibinfo{number}{2} (\bibinfo{year}{2021}), \bibinfo{pages}{119--124}.
\newblock
\href{https://doi.org/10.1038/s41584-020-00539-1}{doi:\nolinkurl{10.1038/s41584-020-00539-1}}


\bibitem[Siu et~al\mbox{.}(2021)]%
        {siu_evaluation_2021}
\bibfield{author}{\bibinfo{person}{Ho~Chit Siu}, \bibinfo{person}{Jaime Peña}, \bibinfo{person}{Edenna Chen}, \bibinfo{person}{Yutai Zhou}, \bibinfo{person}{Victor Lopez}, \bibinfo{person}{Kyle Palko}, \bibinfo{person}{Kimberlee Chang}, {and} \bibinfo{person}{Ross Allen}.} \bibinfo{year}{2021}\natexlab{}.
\newblock \showarticletitle{Evaluation of {Human}-{AI} {Teams} for {Learned} and {Rule}-{Based} {Agents} in {Hanabi}}. In \bibinfo{booktitle}{\emph{Advances in {Neural} {Information} {Processing} {Systems}}}, \bibfield{editor}{\bibinfo{person}{M.~Ranzato}, \bibinfo{person}{A.~Beygelzimer}, \bibinfo{person}{Y.~Dauphin}, \bibinfo{person}{P.~S. Liang}, {and} \bibinfo{person}{J.~Wortman Vaughan}} (Eds.), Vol.~\bibinfo{volume}{34}. \bibinfo{publisher}{Curran Associates, Inc.}, \bibinfo{pages}{16183--16195}.
\newblock
\urldef\tempurl%
\url{https://proceedings.neurips.cc/paper_files/paper/2021/file/86e8f7ab32cfd12577bc2619bc635690-Paper.pdf}
\showURL{%
\tempurl}


\bibitem[Sobkowicz et~al\mbox{.}(2012)]%
        {SOBKOWICZ2012470}
\bibfield{author}{\bibinfo{person}{Pawel Sobkowicz}, \bibinfo{person}{Michael Kaschesky}, {and} \bibinfo{person}{Guillaume Bouchard}.} \bibinfo{year}{2012}\natexlab{}.
\newblock \showarticletitle{Opinion mining in social media: Modeling, simulating, and forecasting political opinions in the web}.
\newblock \bibinfo{journal}{\emph{Government Information Quarterly}} \bibinfo{volume}{29}, \bibinfo{number}{4} (\bibinfo{year}{2012}), \bibinfo{pages}{470--479}.
\newblock
\showISSN{0740-624X}
\href{https://doi.org/10.1016/j.giq.2012.06.005}{doi:\nolinkurl{10.1016/j.giq.2012.06.005}}
\newblock
\shownote{Social Media in Government - Selections from the 12th Annual International Conference on Digital Government Research (dg.o2011)}.


\bibitem[Thoma et~al\mbox{.}(2018)]%
        {thoma_establishing_2018}
\bibfield{author}{\bibinfo{person}{Brent Thoma}, \bibinfo{person}{Victoria Brazil}, \bibinfo{person}{Jesse Spurr}, \bibinfo{person}{Janice Palaganas}, \bibinfo{person}{Walter Eppich}, \bibinfo{person}{Vincent Grant}, {and} \bibinfo{person}{Adam Cheng}.} \bibinfo{year}{2018}\natexlab{}.
\newblock \showarticletitle{Establishing a {Virtual} {Community} of {Practice} in {Simulation}: {The} {Value} of {Social} {Media}}.
\newblock \bibinfo{journal}{\emph{Simulation in Healthcare}} \bibinfo{volume}{13}, \bibinfo{number}{2} (\bibinfo{date}{April} \bibinfo{year}{2018}), \bibinfo{pages}{124}.
\newblock
\showISSN{1559-2332}
\href{https://doi.org/10.1097/SIH.0000000000000284}{doi:\nolinkurl{10.1097/SIH.0000000000000284}}


\bibitem[Törnberg et~al\mbox{.}(2023)]%
        {törnberg2023simulating}
\bibfield{author}{\bibinfo{person}{Petter Törnberg}, \bibinfo{person}{Diliara Valeeva}, \bibinfo{person}{Justus Uitermark}, {and} \bibinfo{person}{Christopher Bail}.} \bibinfo{year}{2023}\natexlab{}.
\newblock \bibinfo{title}{Simulating Social Media Using Large Language Models to Evaluate Alternative News Feed Algorithms}.
\newblock
\showeprint[arxiv]{2310.05984}~[cs.SI]


\bibitem[Vinyals et~al\mbox{.}(2019)]%
        {vinyals_grandmaster_2019}
\bibfield{author}{\bibinfo{person}{Oriol Vinyals}, \bibinfo{person}{Igor Babuschkin}, \bibinfo{person}{Wojciech~M. Czarnecki}, \bibinfo{person}{Michaël Mathieu}, \bibinfo{person}{Andrew Dudzik}, \bibinfo{person}{Junyoung Chung}, \bibinfo{person}{David~H. Choi}, \bibinfo{person}{Richard Powell}, \bibinfo{person}{Timo Ewalds}, \bibinfo{person}{Petko Georgiev}, \bibinfo{person}{Junhyuk Oh}, \bibinfo{person}{Dan Horgan}, \bibinfo{person}{Manuel Kroiss}, \bibinfo{person}{Ivo Danihelka}, \bibinfo{person}{Aja Huang}, \bibinfo{person}{Laurent Sifre}, \bibinfo{person}{Trevor Cai}, \bibinfo{person}{John~P. Agapiou}, \bibinfo{person}{Max Jaderberg}, \bibinfo{person}{Alexander~S. Vezhnevets}, \bibinfo{person}{Rémi Leblond}, \bibinfo{person}{Tobias Pohlen}, \bibinfo{person}{Valentin Dalibard}, \bibinfo{person}{David Budden}, \bibinfo{person}{Yury Sulsky}, \bibinfo{person}{James Molloy}, \bibinfo{person}{Tom~L. Paine}, \bibinfo{person}{Caglar Gulcehre}, \bibinfo{person}{Ziyu Wang}, \bibinfo{person}{Tobias Pfaff},
  \bibinfo{person}{Yuhuai Wu}, \bibinfo{person}{Roman Ring}, \bibinfo{person}{Dani Yogatama}, \bibinfo{person}{Dario Wünsch}, \bibinfo{person}{Katrina McKinney}, \bibinfo{person}{Oliver Smith}, \bibinfo{person}{Tom Schaul}, \bibinfo{person}{Timothy Lillicrap}, \bibinfo{person}{Koray Kavukcuoglu}, \bibinfo{person}{Demis Hassabis}, \bibinfo{person}{Chris Apps}, {and} \bibinfo{person}{David Silver}.} \bibinfo{year}{2019}\natexlab{}.
\newblock \showarticletitle{Grandmaster level in {StarCraft} {II} using multi-agent reinforcement learning}.
\newblock \bibinfo{journal}{\emph{Nature}} \bibinfo{volume}{575}, \bibinfo{number}{7782} (\bibinfo{date}{Nov.} \bibinfo{year}{2019}), \bibinfo{pages}{350--354}.
\newblock
\showISSN{1476-4687}
\href{https://doi.org/10.1038/s41586-019-1724-z}{doi:\nolinkurl{10.1038/s41586-019-1724-z}}
\newblock
\shownote{Number: 7782 Publisher: Nature Publishing Group}.


\bibitem[Wakefield and Wakefield(2016)]%
        {wakefield_social_2016}
\bibfield{author}{\bibinfo{person}{Robin Wakefield} {and} \bibinfo{person}{Kirk Wakefield}.} \bibinfo{year}{2016}\natexlab{}.
\newblock \showarticletitle{Social media network behavior: {A} study of user passion and affect}.
\newblock \bibinfo{journal}{\emph{The Journal of Strategic Information Systems}} \bibinfo{volume}{25}, \bibinfo{number}{2} (\bibinfo{year}{2016}), \bibinfo{pages}{140--156}.
\newblock
\showISSN{0963-8687}
\href{https://doi.org/10.1016/j.jsis.2016.04.001}{doi:\nolinkurl{10.1016/j.jsis.2016.04.001}}


\bibitem[Wang et~al\mbox{.}(2023)]%
        {wang2023voyager}
\bibfield{author}{\bibinfo{person}{Guanzhi Wang}, \bibinfo{person}{Yuqi Xie}, \bibinfo{person}{Yunfan Jiang}, \bibinfo{person}{Ajay Mandlekar}, \bibinfo{person}{Chaowei Xiao}, \bibinfo{person}{Yuke Zhu}, \bibinfo{person}{Linxi Fan}, {and} \bibinfo{person}{Anima Anandkumar}.} \bibinfo{year}{2023}\natexlab{}.
\newblock \bibinfo{title}{Voyager: An Open-Ended Embodied Agent with Large Language Models}.
\newblock
\showeprint[arxiv]{2305.16291}~[cs.AI]
\urldef\tempurl%
\url{https://arxiv.org/abs/2305.16291}
\showURL{%
\tempurl}


\bibitem[Wei and Nguyen(2019)]%
        {9014365}
\bibfield{author}{\bibinfo{person}{Feng Wei} {and} \bibinfo{person}{Uyen~Trang Nguyen}.} \bibinfo{year}{2019}\natexlab{}.
\newblock \showarticletitle{Twitter Bot Detection Using Bidirectional Long Short-Term Memory Neural Networks and Word Embeddings}. In \bibinfo{booktitle}{\emph{2019 First IEEE International Conference on Trust, Privacy and Security in Intelligent Systems and Applications (TPS-ISA)}}. \bibinfo{pages}{101--109}.
\newblock
\href{https://doi.org/10.1109/TPS-ISA48467.2019.00021}{doi:\nolinkurl{10.1109/TPS-ISA48467.2019.00021}}


\bibitem[Wei et~al\mbox{.}(2022)]%
        {NEURIPS2022_9d560961}
\bibfield{author}{\bibinfo{person}{Jason Wei}, \bibinfo{person}{Xuezhi Wang}, \bibinfo{person}{Dale Schuurmans}, \bibinfo{person}{Maarten Bosma}, \bibinfo{person}{brian ichter}, \bibinfo{person}{Fei Xia}, \bibinfo{person}{Ed Chi}, \bibinfo{person}{Quoc~V Le}, {and} \bibinfo{person}{Denny Zhou}.} \bibinfo{year}{2022}\natexlab{}.
\newblock \showarticletitle{Chain-of-Thought Prompting Elicits Reasoning in Large Language Models}. In \bibinfo{booktitle}{\emph{Advances in Neural Information Processing Systems}}, \bibfield{editor}{\bibinfo{person}{S.~Koyejo}, \bibinfo{person}{S.~Mohamed}, \bibinfo{person}{A.~Agarwal}, \bibinfo{person}{D.~Belgrave}, \bibinfo{person}{K.~Cho}, {and} \bibinfo{person}{A.~Oh}} (Eds.), Vol.~\bibinfo{volume}{35}. \bibinfo{publisher}{Curran Associates, Inc.}, \bibinfo{pages}{24824--24837}.
\newblock
\urldef\tempurl%
\url{https://proceedings.neurips.cc/paper_files/paper/2022/file/9d5609613524ecf4f15af0f7b31abca4-Paper-Conference.pdf}
\showURL{%
\tempurl}


\bibitem[Wooldridge and Jennings(1995)]%
        {wooldridge_jennings_1995}
\bibfield{author}{\bibinfo{person}{Michael Wooldridge} {and} \bibinfo{person}{Nicholas~R. Jennings}.} \bibinfo{year}{1995}\natexlab{}.
\newblock \showarticletitle{Intelligent agents: theory and practice}.
\newblock \bibinfo{journal}{\emph{The Knowledge Engineering Review}} \bibinfo{volume}{10}, \bibinfo{number}{2} (\bibinfo{year}{1995}), \bibinfo{pages}{115–152}.
\newblock
\href{https://doi.org/10.1017/S0269888900008122}{doi:\nolinkurl{10.1017/S0269888900008122}}


\bibitem[Wu et~al\mbox{.}(2022)]%
        {10.1145/3491101.3519729}
\bibfield{author}{\bibinfo{person}{Tongshuang Wu}, \bibinfo{person}{Ellen Jiang}, \bibinfo{person}{Aaron Donsbach}, \bibinfo{person}{Jeff Gray}, \bibinfo{person}{Alejandra Molina}, \bibinfo{person}{Michael Terry}, {and} \bibinfo{person}{Carrie~J Cai}.} \bibinfo{year}{2022}\natexlab{}.
\newblock \showarticletitle{PromptChainer: Chaining Large Language Model Prompts through Visual Programming}. In \bibinfo{booktitle}{\emph{Extended Abstracts of the 2022 CHI Conference on Human Factors in Computing Systems}} (New Orleans, LA, USA) \emph{(\bibinfo{series}{CHI EA '22})}. \bibinfo{publisher}{Association for Computing Machinery}, \bibinfo{address}{New York, NY, USA}, Article \bibinfo{articleno}{359}, \bibinfo{numpages}{10}~pages.
\newblock
\showISBNx{9781450391566}
\href{https://doi.org/10.1145/3491101.3519729}{doi:\nolinkurl{10.1145/3491101.3519729}}


\bibitem[Xi et~al\mbox{.}(2023)]%
        {xi2023rise}
\bibfield{author}{\bibinfo{person}{Zhiheng Xi}, \bibinfo{person}{Wenxiang Chen}, \bibinfo{person}{Xin Guo}, \bibinfo{person}{Wei He}, \bibinfo{person}{Yiwen Ding}, \bibinfo{person}{Boyang Hong}, \bibinfo{person}{Ming Zhang}, \bibinfo{person}{Junzhe Wang}, \bibinfo{person}{Senjie Jin}, \bibinfo{person}{Enyu Zhou}, \bibinfo{person}{Rui Zheng}, \bibinfo{person}{Xiaoran Fan}, \bibinfo{person}{Xiao Wang}, \bibinfo{person}{Limao Xiong}, \bibinfo{person}{Yuhao Zhou}, \bibinfo{person}{Weiran Wang}, \bibinfo{person}{Changhao Jiang}, \bibinfo{person}{Yicheng Zou}, \bibinfo{person}{Xiangyang Liu}, \bibinfo{person}{Zhangyue Yin}, \bibinfo{person}{Shihan Dou}, \bibinfo{person}{Rongxiang Weng}, \bibinfo{person}{Wensen Cheng}, \bibinfo{person}{Qi Zhang}, \bibinfo{person}{Wenjuan Qin}, \bibinfo{person}{Yongyan Zheng}, \bibinfo{person}{Xipeng Qiu}, \bibinfo{person}{Xuanjing Huang}, {and} \bibinfo{person}{Tao Gui}.} \bibinfo{year}{2023}\natexlab{}.
\newblock \bibinfo{title}{The Rise and Potential of Large Language Model Based Agents: A Survey}.
\newblock
\showeprint[arxiv]{2309.07864}~[cs.AI]


\bibitem[Yu et~al\mbox{.}(2023)]%
        {10144391}
\bibfield{author}{\bibinfo{person}{Junliang Yu}, \bibinfo{person}{Hongzhi Yin}, \bibinfo{person}{Xin Xia}, \bibinfo{person}{Tong Chen}, \bibinfo{person}{Jundong Li}, {and} \bibinfo{person}{Zi Huang}.} \bibinfo{year}{2023}\natexlab{}.
\newblock \showarticletitle{Self-Supervised Learning for Recommender Systems: A Survey}.
\newblock \bibinfo{journal}{\emph{IEEE Transactions on Knowledge and Data Engineering}} (\bibinfo{year}{2023}), \bibinfo{pages}{1--20}.
\newblock
\href{https://doi.org/10.1109/TKDE.2023.3282907}{doi:\nolinkurl{10.1109/TKDE.2023.3282907}}


\bibitem[Zanettin({[n.\,d.]})]%
        {twitter}
\bibfield{author}{\bibinfo{person}{Federico Zanettin}.} \bibinfo{year}{[n.\,d.]}\natexlab{}.
\newblock \bibinfo{title}{X Develop Platform}.
\newblock
\urldef\tempurl%
\url{https://developer.x.com/en/docs/x-api/getting-started/about-x-api#item0}
\showURL{%
\tempurl}


\bibitem[Zimmer(2010)]%
        {10.1007/s10676-010-9227-5}
\bibfield{author}{\bibinfo{person}{Michael Zimmer}.} \bibinfo{year}{2010}\natexlab{}.
\newblock \showarticletitle{"But the data is already public": on the ethics of research in Facebook}.
\newblock \bibinfo{journal}{\emph{Ethics and Inf. Technol.}} \bibinfo{volume}{12}, \bibinfo{number}{4} (\bibinfo{date}{Dec.} \bibinfo{year}{2010}), \bibinfo{pages}{313–325}.
\newblock
\showISSN{1388-1957}
\href{https://doi.org/10.1007/s10676-010-9227-5}{doi:\nolinkurl{10.1007/s10676-010-9227-5}}


\end{thebibliography}
\appendix
\section{Appendix}
\subsection{Social Media Engine}
The core of the prompt template is shown below:
\begin{verbatim}
    There is basic information of {Agent1}. {Agent1's Demographic Information}
    {Agent2} posted that {Content} at {Time}.
    On a scale of 1 to 10, where 1 is not recommended and 10 is highly recommended,
    rate the likely that Sparkle recommend the {Agent2}'s post to {Agent1}.
    Rate (return a number between 1 to 10):
\end{verbatim}
The content within “\{\}” can be modified based on specific circumstances.
Additionally, users have the option to customize the recommendation threshold.

\subsection{Cognitive Architecture}
\subsubsection{Importance Measure}\label{important}
\re{The prompt template is shown below:}
\begin{verbatim}
    Here is a brief description of {Agent}. 
    {Agent's Demographic Information}
    On the scale of 1 to 10, where 1 is purely mundane (e.g., brushing teeth, making bed) 
    and 10 is extremely poignant (e.g., a break up, college acceptance), 
    rate the likely poignancy of the following event for {Agent}.
    Event: {Event}
    Rate (return a number between 1 to 10):
\end{verbatim}
\subsubsection{Decide Module}\label{decide}
We provide the core of a prompt example of whether the agent decides to post on social media:
\begin{verbatim}
    Context: {Perceptions} 
    Right now, it is {Time}. {Agent} last posted about {Content} at {Time}. 
    {Agent's Demographic Information}
    {Agent} usually around {Post Frequency}.
    {Agent's Retrieved Memories}
    Question: Would {Agent} post right now? 
    Reasoning: Let's think step by step.
    Output the response to the prompt above.
    The output should contain two parts (Reasoning and Answer). 
    The output should be in JSON.
\end{verbatim}
Given the prompt above, \tocheck{the reasoning and decision will be provided together. } The possible output that the agent would not post will be:
\begin{verbatim}
    {"Reasoning":"Rebecca is a data analyst who loves exploring data trends.
    She usually posts around 1 times a day, and has already posted once today.
    Therefore, it is unlikely that she would post again right now",
    "Answer":"No"}
\end{verbatim}

\subsubsection{Acting}\label{acting}
Assuming that an agent has decided to post, the corresponding prompt core in the acting part will be:
\begin{verbatim}
    Context: {Perceptions} 
    Right now, it is {Time}. {Agent} last posted about {Content} at {Time}. 
    {Agent's Demographic Information}
    {Agent's Retrieved Memories}
    Question: What would {Agent} post right now? 
    Output the response to the prompt above.
    The output should contain post Content. 
    The output should be in JSON
\end{verbatim}

\subsubsection{Chain-of-thought Strategy}\label{achain}
\re{Each prompt basically consists of four parts including the \textbf{core contents},} \textbf{particular statements} ``Let's think step by step.'', \textbf{specific instructions}\re{~(\eg ``The output should be in JSON'')} and \textbf{output examples}~(R1).

\subsection{User Study}\label{auser}
\re{We followed chain-of-thought rules mentioned in \Cref{chain} to design prompt.
The core part of the prompt is shown below:}
\begin{verbatim}
    Please generate 20 different agents' profiles, including name, age, 
    gender, residency, innate, job, and lifestyle,
    together with their social media habits including followers, posting frequency,
    primary content of posts, and engagements.
\end{verbatim}
\re{As for specific instructions:}
\begin{verbatim}
    The profiles of the agents should resemble real humans, 
    and the differences between each agent should be maximized. 
    The final output should be in JSON format.
\end{verbatim}
\end{document}